\documentclass[%
 aps,
  pre,
  superscriptaddress, letterpaper,
  longbibliography,
  twocolumn,
  showpacs,
  showkeys,
  amsfonts, amssymb, amsmath]{revtex4-1}

\usepackage{xcolor}
\usepackage[applemac]{inputenc}
\usepackage{graphicx,amssymb,amsmath,array,multirow}
\usepackage{hyperref} 
\hypersetup{pdfpagemode={UseOutlines},
bookmarksopen=true,
bookmarksopenlevel=0,
hypertexnames=false,
colorlinks=true, 
citecolor=blue, 
linkcolor=red, 
urlcolor=black, 
pdfstartview={FitV},
unicode,
breaklinks=true,
}

\definecolor{pink}{rgb}{1,1,0} 
\definecolor{red}{rgb}{1,0,0}
\definecolor{yellow}{rgb}{1,1,0}
\definecolor{orange}{rgb}{1,0.5,0} 
\definecolor{blue}{rgb}{0,0,1}
\definecolor{white}{rgb}{1,1,1}

\def\NOTE#1{{\textcolor{red}{\bf [#1]}}}   
\def\NOTE2#1{{\textcolor{blue}{\bf [#1]}}}   

\newcommand{\spcode}{\textsc{spc}}
\newcommand{\mcode}{\textsc{mc}}

\begin{document}

\title{Elastic Weak Turbulence : from the vibrating plate to the drum}
\author{Roumaissa Hassaini}
\author{Nicolas Mordant}
\email[]{nicolas.mordant@univ-grenoble-alpes.fr}
\affiliation{Laboratoire des Ecoulements Geophysiques et Industriels, Universite Grenoble Alpes, CNRS, Grenoble-INP,  F-38000 Grenoble, France}
\author{Benjamin Miquel}
\affiliation{Department of Applied Mathematics, University of Colorado, Boulder}
\author{Giorgio Krstulovic}
\affiliation{Universit\'e C\^ote d'Azur, Observatoire de la C\^ote d'Azur, CNRS, Laboratoire Lagrange, Bd de l'Observatoire, CS 34229, 06304 Nice cedex 4, France.}
\author{Gustavo D\"uring}
\email[]{gduring@fisica.puc.cl}
\affiliation{Instituto de F\'isica, Pontificia Universidad Cat\'olica de Chile, Casilla 306, Santiago, Chile.}

\begin{abstract}
Weak wave turbulence has been observed on a thin elastic plate in previous work. Here we report theoretical, experimental and numerical studies of wave turbulence in a thin elastic plate submitted to increasing tension. When increasing the tension (or decreasing the bending stiffness of the plate) the plate evolves progressively from a plate into an elastic membrane as in drums. We start from the plate and increase the tension in experiments and numerical simulations. We observe that the system remains in a state of weak turbulence of weakly dispersive waves. This observation is in contrast with what has been observed in water waves when decreasing the water depth, which also changes the waves from dispersive to weakly dispersive. The weak turbulence observed in the deep water case evolves into a solitonic regime. Here no such transition is observed for the stretched plate. We then apply the weak turbulence theory to the membrane case and show with numerical simulations that indeed the weak turbulence framework remains valid for the membrane and no formation of singular structures (shocks) should be expected in contrast with acoustic wave turbulence. 
\end{abstract}

\maketitle

\section{Introduction}

Wave turbulence is a generic class of systems in which a large number of waves coupled through nonlinearity evolve into a complex statistical state. The most natural physical system in which wave turbulence occurs is that of water waves at the surface of the ocean: Waves are forced by the wind and once their amplitude is large enough they develop a wideband spectrum. Depending on the strength of nonlinearity, various scaling behaviors of the spectrum have been reported (see for instance recent field measurements in \cite{Leckler,Lenain}). In the limit of weak nonlinearity a statistical theory named Weak Turbulence Theory (WTT) has been developed by Hasselman~\cite{Hasselmann} that allows to predict the evolution equation of the wave spectrum and to exhibit its stationary solutions (following works of Zakharov in particular~\cite{R1,Nazarenko,newell_wave_2011}). The WTT has been applied to a large number of other systems such as inertial waves in astrophysical or geophysical flows, magnetized plasmas, optics in non linear media, superfluid turbulence, among others \cite{Galtier,Scott,R1,laurie2012one,R16}. In all these systems waves are dispersive which means that waves at different frequencies propagate at different velocities. As in the WTT the non linear coupling is assumed to occur through N-wave resonances ($N\geq 3$), the dispersive character of the waves enables non trivial solutions of the resonance equations such as
\begin{equation}
\omega_1=\omega_2+\omega_3, \quad \quad \mathbf k_1=\mathbf k_2+\mathbf k_3
\end{equation}
in the case $N=3$ ($\omega_i$ are angular frequencies and $\mathbf k_i$ are wavevectors). The question of what may happen for non dispersive waves is largely open and depends most likely on the details of the nonlinear coupling. Indeed for non dispersive waves, the only resonant solutions for 3-wave coupling is through waves propagating in the same direction. Thus the waves may develop shocks by a cumulative effect of the nonlinearity as is the case for acoustic waves for instance \cite{aucoin,acustictur}. In the case of water waves at the surface of shallow water, when the waves are both weakly dispersive and weakly nonlinear, solitons can be observed rather than the interplay of free dispersive waves that is expected in the WTT. When reducing the depth of water, a transition from WTT to a solitonic regime could be observed experimentally~\cite{hassaini2017transition}. A question is whether a similar transition would occur for other systems. Weak turbulence was predicted for instance for gravitational waves in the early universe~\cite{Galtier:2017hi}. In that case waves are non dispersive and the coupling occurs through 4-wave interaction. A regime of weak turbulence is predicted nonetheless although no validation through numerical simulations (let alone experiments !) has been reported to our knowledge. 

Here we focus on a distinct physical system which provides a paradigm for studying the transition non dispersive wave turbulence by means of both experimental realizations and computational studies: the elastic plate under tension. The elastic plate is a rich physical system for wave turbulence which has been the object of a growing interest~\cite{during,Boudaoud,R19,Yokoyama1,Yokoyama2,Humbert,Ducceschi,platesPhysD,Mordant:2017bh,During:2018jc}. A shaken metal plate has been used for centuries to mimic the thunder noise in theaters~\cite{R17} and in the first half of the 20th century as a reverberator for analog signal processing~\cite{arcasICA07}. Beyond these somewhat specialized applications, the vibrating plate has become recently a privileged model for wave turbulence. Indeed, advanced measurement resolved in both space and time are possible experimentally~\cite{R19} and its 2D character facilitates long numerical simulations~\cite{during,miquel_role_2014,Ducceschi} (by contrast with  substantially more challenging 3D fluid systems, for instance).
It was used to validate numerically the WTT prediction~\cite{during,miquel_role_2014}, to investigate the effect of wideband dissipation~\cite{miquel_role_2014,Humbert}, the impact of finite size~\cite{epjb} or the strongly nonlinear regime~\cite{Yokoyama1,Yokoyama2,miquel2013transition,Mordant:2017bh}. Thereafter, we consider the addition of tension to elastic plates plate as a tuning parameter for the dispersive character of the waves. When tension is increased to the point that it dominates over bending stiffness as a restoring force for wave motions, the plate starts to behave like a membrane for which the waves are non dispersive.

A thin elastic plate without tension can be modeled by the F\"oppl-von Karman equations
\cite{foppl,vk,landau} for the amplitude of the deformation $\zeta(x,y,t)$ of the plate and for  the Airy stress function $\chi(x,y,t)$:
\begin{eqnarray} 
\rho\frac{\partial^2 \zeta}{\partial t^2} &=& - \frac{Eh^2}{12(1-\mu^2)}\Delta^2\zeta +
\{\zeta,\chi\}  ;
\label{foppl0}\\
\Delta^2\chi &=&- \frac{E}{2}\{\zeta,\zeta\},
\label{foppl1}
\end{eqnarray}
where  $h$ is the thickness of the elastic sheet. The material has a  mass density $\rho$, a Young modulus $E$ and its Poisson
ratio is $\mu$. $h$ is the thickness of the plate.
$\Delta=\partial_{xx}+\partial_{yy}$ is the usual Laplacian and the bracket $\{\cdot,\cdot\}$ is defined by
$\{f,g\}\equiv f_{xx}g_{yy}+f_{yy}g_{xx}-2f_{xy}g_{xy}$.  Equation (\ref{foppl1}) for  the Airy stress function
$\chi(x,y,t)$ may be seen as the compatibility equation for the in--plane stress tensor: In the derivation the inertia of the in-plane modes of oscillations is neglected, or, in other words, it is assumed that the in-plane displacements are negligible, hence equation (\ref{foppl1}) follows the dynamics. 

An isotropic external tension can be added to the elastic plate considering an isotropic in plane stress given via a boundary condition for the Airy function $\chi(x,y) = \frac{T}{2h}(x^2+y^2),$ where $T$ is the tension per unit length at infinity ($T$ is applied at the boundaries in practice). Then the Airy function can be separated in two terms $\chi \rightarrow \frac{T}{2h}(x^2+y^2) +\chi(x,y) $ in which the first comes from the external tension while the second comes from the amplitude deformation. The F\"oppl--von K\'arm\'an equations (\ref{foppl0}) for a plate under tension then reads
\begin{eqnarray} 
\rho\frac{\partial^2 \zeta}{\partial t^2} &=& - \frac{Eh^2}{12(1-\mu^2)}\Delta^2\zeta +
\{\zeta,\chi\}  + \frac{T}{h} \Delta \zeta ,
\label{fopplTension}
\end{eqnarray}
and equation (\ref{foppl1}) remains unchanged. In the linear limit, waves follow the dispersion relation
\begin{equation}
\omega_{\bf k} =  \left( \frac{h^2E}{12(1-\mu^2)\rho}  k^4 + \frac{T}{\rho h} k^2 \right)^{1/2}.
\label{dispersion.tension}
\end{equation}
This reasoning can be generalized to the case of anisotropic tension, which we study experimentally. Denoting $T_x$ and $T_y$ the tension per unit length at infinity along the $x$ and $y$ direction, respectively, the governing equation (\ref{fopplTension}) yields:
\begin{equation} 
\rho\frac{\partial^2 \zeta}{\partial t^2} = - \frac{Eh^2}{12(1-\mu^2)}\Delta^2\zeta +
\{\zeta,\chi\}  + \frac{1}{h}\left(T_x\partial_{xx}+T_y\partial_{yy}\right) \zeta \, ,
\label{fopplTensionAnisotropic}
\end{equation}
and the corresponding dispersion relation becomes:
\begin{equation}
\omega_{\bf k} =  \left( \frac{h^2E}{12(1-\mu^2)\rho}  k^4 + \frac{1}{\rho h} \left[T_xk_x^2 + T_yk_y^2\right]\right)^{1/2}.
\label{dispersion.tensionAnisotropic}
\end{equation}

For zero tension ($T=0$), i.e. for pure bending, the dispersion relation simplifies into 
\begin{equation}
\omega_{\bf k} =  \sqrt{\frac{h^2E}{12(1-\mu^2)\rho}}  k^2=Ck^2\, ,
\label{rdnotens}
\end{equation}
with $C^2=\frac{h^2E}{12(1-\mu^2)\rho}$.
These waves are dispersive, thus the WTT formalism could be used and yielded predictions for the wave spectra~\cite{during}. These predictions have been verified numerically~\cite{during} but not experimentally due to the presence of wideband dissipation that alters the predicted scaling as dissipation is not taken into account in the theory~\cite{Boudaoud,R19,Humbert,miquel_role_2014}. 

In the opposite limit of no bending stiffness ($E=0$), and in the case of isotropic tension, the dispersion relation changes into
\begin{equation}
\omega_{\bf k} =  \sqrt{\frac{T}{\rho h}} k \,.
\end{equation}
This dispersion relation is now non dispersive but the nonlinear term remains the same. Thus by tuning the value of the tension $T$ and/or the thickness $h$ one can change the system into pure flexion wave or into a membrane without rigidity as in drums and study the evolution of the statistical properties of this new system. 

The crossover between both extreme cases occurs when both contributions are equal, i.e. 
\begin{equation}
k_c=\sqrt{\frac{12 T (1-\mu^2)}{h^3E}} \textrm{ or } \omega_c=\sqrt{\frac{24T^2(1-\mu^2)}{\rho h^4E}}
\label{cross}
\end{equation}
This is somehow analogous to the transition between shallow to deep water regimes for water surface waves that occurs when $kd\approx 1$ ($d$ being the water depth)~\cite{Guyon}.

Our goal is to investigate if the system remains into a regime of weak turbulence when increasing the effect of tension (i.e. when $k_c$ increases). Indeed, the waves being weakly non dispersive to non dispersive the system could evolve into a distinct state involving localized structures such as solitons (as for water waves) or shocks (as for acoustics). Note that we remain in a weakly nonlinear regime so that to avoid the creation of singular structures such as developable cones that have been reported in plates at strong level of nonlinearity~\cite{miquel2013transition,Mordant:2017bh}. Thus we try to answer the question of Newell and Rumpf\cite{newell_wave_2011}: by aiming towards a non dispersive system, will the non-linear interactions redistribute the energy isotropically or will it be focused in shock waves?

In the next section we describe the experiment and the numerical simulations used in this article. In section~\ref{prestre} we first investigate the effect of adding a tension to a plate.As weak wave turbulence has been reported in a vibrating plate, thus we start from such a state and analyse the effect of increasing the tension in a plate either experimentally or numerically. In section~\ref{pure}, we report the application of WTT to the case of a pure membrane with strictly no bending stiffness and finally we use numerical simulations to compare the statistics of a vibrating membrane to the theoretical predictions. 

\section{Methods}

\subsection{Description of the experiment}

\begin{figure}[!htb]
\includegraphics[clip,width=6cm]{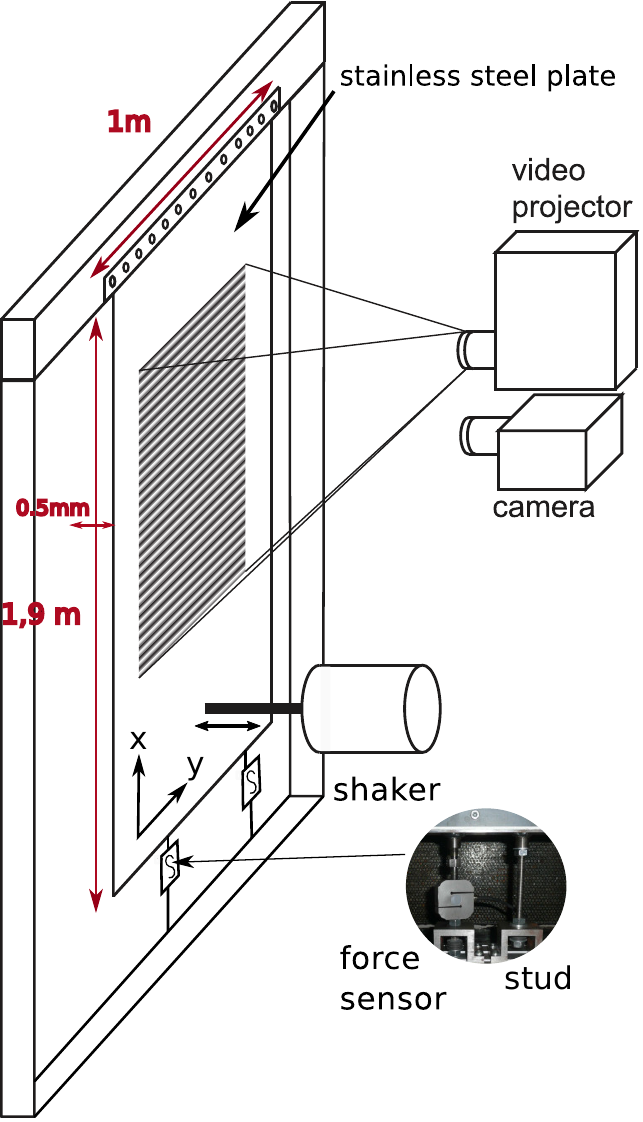}
\caption{Experimental set-up. A $1.9\times 1$~m$^2$, 0.5~mm-thick steel plate is stretched vertically. Vibrations are generated by an electromagnetic shaker at the bottom of the plate. The deformation is measured by the Fourier Transform Profilometry technique in which a pattern is projected on the plate and recorded by a high speed camera (see text for details).}
\label{setup}
\end{figure}
The experimental set-up is composed of a $1\times1.9$~m$^2$ stainless steel plate, $h=0.5$~mm-thick. The physical parameters are $E\approx 2\, 10^{11}$~Pa, $\mu\approx0.3$, $\rho\approx7800$~kg/m$^3$. The plate is attached at its top to a rigid frame (fig.~\ref{setup}). Tension can be added by using left-right threaded studs at the bottom. The tension is thus only along the long axis of the plate for reasons of simplicity: in other words, we may vary the tension along the vertical direction $T_x$ while the transverse tension $T_y$ stays zero. The tension $T_x$ is tuned by first stretching the plate to some level given by two force gauges. A second set of studs is then tightened so that to vanish exactly the tension from the force gauges. The gauges are then uncoupled from the plate to preserve them from the strong vibrations of the plate during experiments. Vibrations are generated by an electromagnetic shaker oscillating sinusoidally at a 30~Hz frequency in the direction normal to the plate and bolted to the plate 30 cm above its bottom.

Fourier transform profilometry is used to measure the deformation of the plate with both space and time resolution following the scheme developed by Cobelli et al.~\cite{cobelli_space-time_2009}.
It consists in projecting a sinusoidal grayscale pattern with a video-projector on a $0.8\times1.40$~m$^2$ region centered in the plate (painted in white). When the waves propagate, the initial pattern is deformed. This distortion is recorded by a high speed camera at a frequency of 8000 frames/s and with a resolution of $704\times1024$ pixels$^{2}$.
A geometrical analysis permits a phase inversion of the images and the reconstruction of the wave field~\cite{Cobelli1,R22}.
The images are recorded once a statistically stationary regime is reached which takes only a few seconds.
Experiments are repeated to increase the amount of data and to achieve a good convergence in the statistical properties measured. 

\subsection{Numerical simulations}
\label{DNS}

We complement the experimental measurements with numerical solutions to the augmented F\"oppl-Von Karman equations (aFVKE). We write down the projection of Eq.~(\ref{fopplTensionAnisotropic}) in Fourier space for a wave vector $\mathbf{k}$:
\begin{equation}
    \partial_{tt}\zeta_{\mathbf k} = -C^{2}k^{4}\zeta_{\mathbf k} - (T_xk_x^2+T_yk_y^2)\zeta_{\mathbf k} + \frac{1}{\rho}\{\zeta,\chi\}_{\mathbf k} + \mathcal F_{\mathbf k} + \mathcal D_{\mathbf k}
    \label{FvK_fourier}
\end{equation}
where $\mathcal{F}_{\mathbf k}$ and $\mathcal{D}_{\mathbf k}$ are the forcing and dissipative terms in spectral space, respectively. We recall the reader that we allow for anisotropic tension ($T_x\neq T_y$) which pertains to the experimental setup. This equation is studied by means of two distinct pseudo-spectral algorithms, documented in this section, depending on the presence of rigidity (i.e., the case of stretched plates with finite $k_c$) or its absence (i.e., the case of membranes with $k_c\rightarrow\infty$). Both the Stretched Plate Code (\spcode~below) and the Membrane Code (\mcode~below) employ a Fourier decomposition of the variables $\zeta$ and $\chi$ and implement periodic boundary conditions, which differ from the experimental boundary conditions. However, the impact of boundary conditions on the wave state has been shown to be marginal~\cite{miquel_role_2014}, provided that the size of the domain is significantly larger than the forcing wavelength. We thus have confidence in the ability of the simulations reported thereafter to capture the meaningful statistical properties of the experimental wave field. Both codes evaluate the non linear terms in physical space, and fields are fully dealiased. We present below the particulars of the \spcode~and \mcode~codes which differ in their time-stepping method, and in the forcing and dissipation employed.

\subsubsection{The Stretched Plate code}

The Stretched Plate Code (\spcode) is an evolution of the code previously used in~\cite{miquel2013transition,miquel_role_2014,miquel_nonlinear_2011}. The timestepping of Eq.~\ref{FvK_fourier} is done by means of the second order exponential time-differencing ETD2RK scheme of Cox and Matthews \cite{cox2002exponential}. The code has been developed in view of providing a way to bridge the gap between imperfect experimental realizations of wave turbulence and the ideal hypotheses of the weak turbulence theory. 

For instance, the code implements the possibility to modify independently the tension along the two directions $T_x$ and $T_y$. This offers the possibility to reproduce the anisotropic tension present in the experiments where the transverse tension vanishes: $T_y=0$.   

A similar approach is adopted with dissipation. As reported by Miquel and Mordant~\cite{miquel_nonlinear_2011}, the experimental dissipation can be modelled with a linear damping $\mathcal D_{\mathbf k}=- \gamma_{k} \partial_t\zeta_{\mathbf k}$, where $\gamma_{k}$ has the analytic form:
\begin{equation}
    \gamma_{k}^{exp} = \gamma_{1} + \gamma_{2}k^{2}\,. \label{eq:gamma_exp}
\end{equation}
In absence of tension, Miquel et al.~\cite{miquel_role_2014} reported that this expression of the dissipation, where the parameters $\gamma_{1}$ and $\gamma_{2}$ are obtained by fitting the experimental data, could reproduce qualitatively the experimental spectra of elastic wave turbulence. Thereafter, we refer to this case as the ``experimental dissipation". This dissipation is much different from the ideal dissipation considered in the weak turbulence theory, which assumes the existence of a transparency window of wave numbers where damping vanishes and, consequently, energy is conservatively transferred. Indeed, it has been shown in~\cite{miquel_role_2014} that the theoretical scaling predicted by the theory is recovered when the dissipation rate is set to zero within a wave number interval $[k_l,k_h]$. In the following, ``ideal dissipation'' refers to a damping coefficient following Eq.~\ref{eq:gamma_exp}, except between $k_{l} = 5$~m$^{-1}$ and $k_{h}=240$~m$^{-1}$ where it is set to zero.

We adopt a forcing that differs from the spatially localized, harmonic, experimental forcing. We force resonantly a crown of modes $\mathbf{k}$ at their linear frequency $\omega_{k}$: 
\begin{equation}
    \mathcal F_{\mathbf k}(t) = F_{0}e^{-\frac{k-k_{0}}{2\nu_{k}^{2}}}e^{i\varphi_{k}}cos(\omega_{k}t + \phi_{k})\, .
\end{equation} The crown of forced wave numbers is centered around $k_{0}=5\pi$~m$^{-1}$, with a width $\sigma_{k}=2\pi $~m$^{-1}$. Random phases $\phi_{k}$ and $\varphi_{k}$ are introduced to obtain a disordered forcing.

 The spatial resolution is moderate ($288^2$) so that the simulations can be run to obtain long time series, up to the equivalent of 300 seconds of experiments for the longest ones. As in experiments, the deformation and the vertical velocity are recorded in time for subsequent spectral analysis.

\subsubsection{The Membrane code}

We complement the previous numerical simulation with simulations at higher resolution from a second pseudo-spectral code, parallelized using GPU computing. We integrate the F\"oppl--von K\'arm\'an equations~(\ref{fopplTensionAnisotropic}) in their dimensionless form and in the limit of pure tension (E=0) which corresponds to a non dispersive membrane. By contrast with the Stretched Plane code described above, the Membrane code implements a random forcing and some hyperviscous dissipation. In Fourier space, Eq.~(\ref{FvK_fourier}) becomes:
\begin{equation}
    \partial_{tt}\zeta_{\mathbf k} =  - \frac{1}{4}k^2\zeta_{\mathbf k} + \frac{1}{\rho}\{\zeta,\chi\}_{\mathbf k} + \mathcal F_{\mathbf k} -\nu_0 k^\alpha \partial_t \zeta_\mathbf{k}
    \label{FvK_fourier_dimless}
    \end{equation}
where $\nu_0$ and $\mathcal{F}_\mathbf{k}(t)$ are the rescaled damping coefficient and rescaled external forcing, respectively.
We supply the system with periodic boundary conditions in a square domain of size $2\pi$. The forcing $\mathcal F_{\mathbf k}$ is white-noise in time of variance $f_0^2$ and its Fourier modes are non-zero only for wave-vectors $k_i<|{\bf k}|\le k_f$. A standard second-order Runge-Kutta time-stepping scheme is implemented. De-aliasing is made by using the standard $2/3$-rule \cite{gottlieb1977numerical}, that is applied after computing each quadratic term. The largest wavenumber is $k_{\rm max}=N/3$, where $N$ is the resolution. In numerics we set an hyperviscous dissipation with $\alpha=6$. Two resolutions were used; $N=1024^{2}$ and $N=512^{2}$.

The main parameters of experiments and numerical simulations are given in table~\ref{tableau}.

\begin{table*}[!htb]
\center
\begin{tabular}{c|c|c|c|c|c|c|c|m{1cm}}
description & name & $C^2$ & $T$(kN/m) & $\omega_c$ & $k_c$ & $N$ & dissipation & forcing scale\\
  \hline
    exp. stretched plate  & E1 &$C^2_{exp}$ & $T_x=4$, $T_y=0$ & 290~Hz, $10\,\omega_f$ & $7k_f$ &   & &large\\
    exp. stretched plate  & E2 &$C^2_{exp}$ & $T_x=10$, $T_y=0$ & 725~Hz, $25\,\omega_f$ & $17k_f$ &   & &large\\
    DNS stretched plate & N1 &$C^2_{exp}$ & $T_x=4$, $T_y=0$  & 290~Hz, $3.5\,\omega_f$ &$2.6k_f$ & $288^2$ & experimental & large\\
    DNS stretched plate & N2 &$C^2_{exp}$ & 4 & $3.5\,\omega_f$ &$2.6k_f$ & $288^2$ & ideal  & large\\
    DNS stretched plate & N3 &$C^2_{exp}/2$ & 4 & $5\,\omega_f$ & $3.7k_f$& $288^2$ & ideal  & large\\
    DNS stretched plate & N4 &$C^2_{exp}/5$ & 4 & $8\,\omega_f$ & $5.8k_f$& $288^2$ & ideal  & large\\
    DNS stretched plate & N5 &$C^2_{exp}/10$ & 4 & $12\,\omega_f$ & $8.3k_f$ & $288^2$ & ideal  & large\\
    DNS stretched plate & N6 &$C^2_{exp}/100$ & 4 & $37\,\omega_f$ & $26k_f>k_d$& $288^2$ & ideal  & large\\
    DNS stretched plate & N7 &$C^2_{exp}/1000$ & 4 & $117\,\omega_f$ & $83k_f>k_d$&$288^2$ & ideal  & large\\
    DNS membrane & N8 &$0$ &  &  & & $288^2$ & ideal  & large\\
    DNS membrane & N9 &$0$ &  & & & $1024^2$ & hypervisc.  & small\\
    DNS membrane & N10 &$0$ &  & & & $512^2$ & hypervisc.  & small\\
    DNS membrane & N11 &$0$ &  &  & & $288^2$ & ideal  & small\\
\end{tabular}
\caption{Description of the various experiments (exp) and numerical simulations (DNS). $C^2$ is such that the dispersion relation is $\omega_{\bf k} =  \left( C^2  k^4 + \frac{T}{\rho h} k^2 \right)^{1/2}$. It corresponds to the dispersive part. The experimental value has been estimated to be $C^2_{exp}=0.6084$ from the observed dispersion relation. $T$ is the tension per unit length (when not specified $T_x=T_y$). $\omega_c$ and $k_c$ are the crossover frequency and wavenumber between dispersive and non dispersive waves (eq.~(\ref{cross})). They are expressed as a function of the forcing frequency $\omega_f$ and forcing wavenumber $k_f=5\pi$~rad/m repectively. $N$ is the number of grid point in numerical simulations. ``dissipation'' is the dissipative scheme that is used in the DNS (see part \ref{DNS}). $k_d=240$~rad/m is the dissipative cutoff of the ``ideal'' dissipation scheme. ``forcing'' specifies whether the forcing acts at large scale to investigate the direct cascade or at small scale to study the inverse cascade.}
\label{tableau}
\end{table*}

\section{Investigation of the stretched plate}
\label{prestre}

\subsection{Experimental results}
\begin{figure*}[!htb]
\includegraphics[clip,width=8cm]{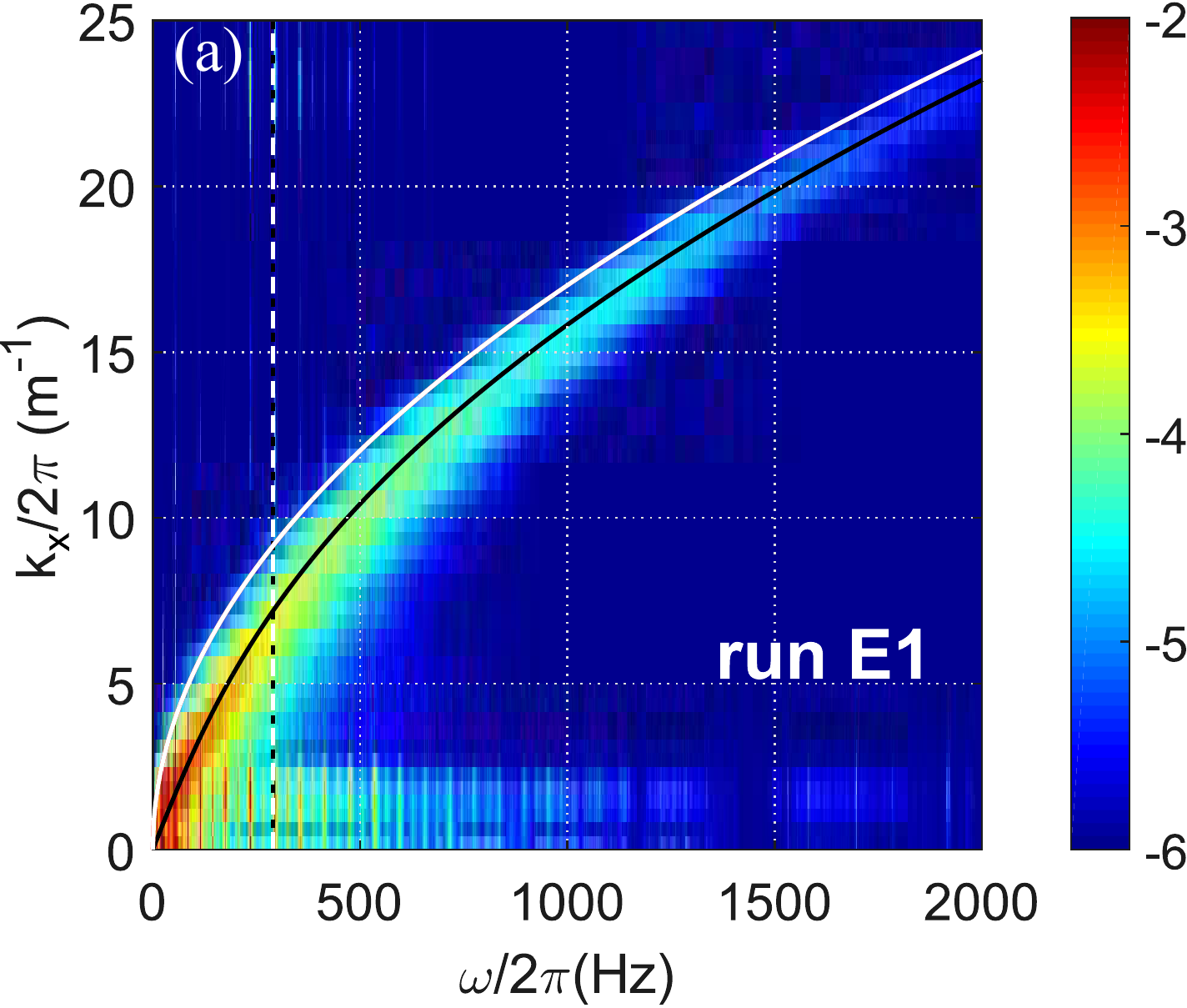}\hfill
 \includegraphics[clip,width=8cm]{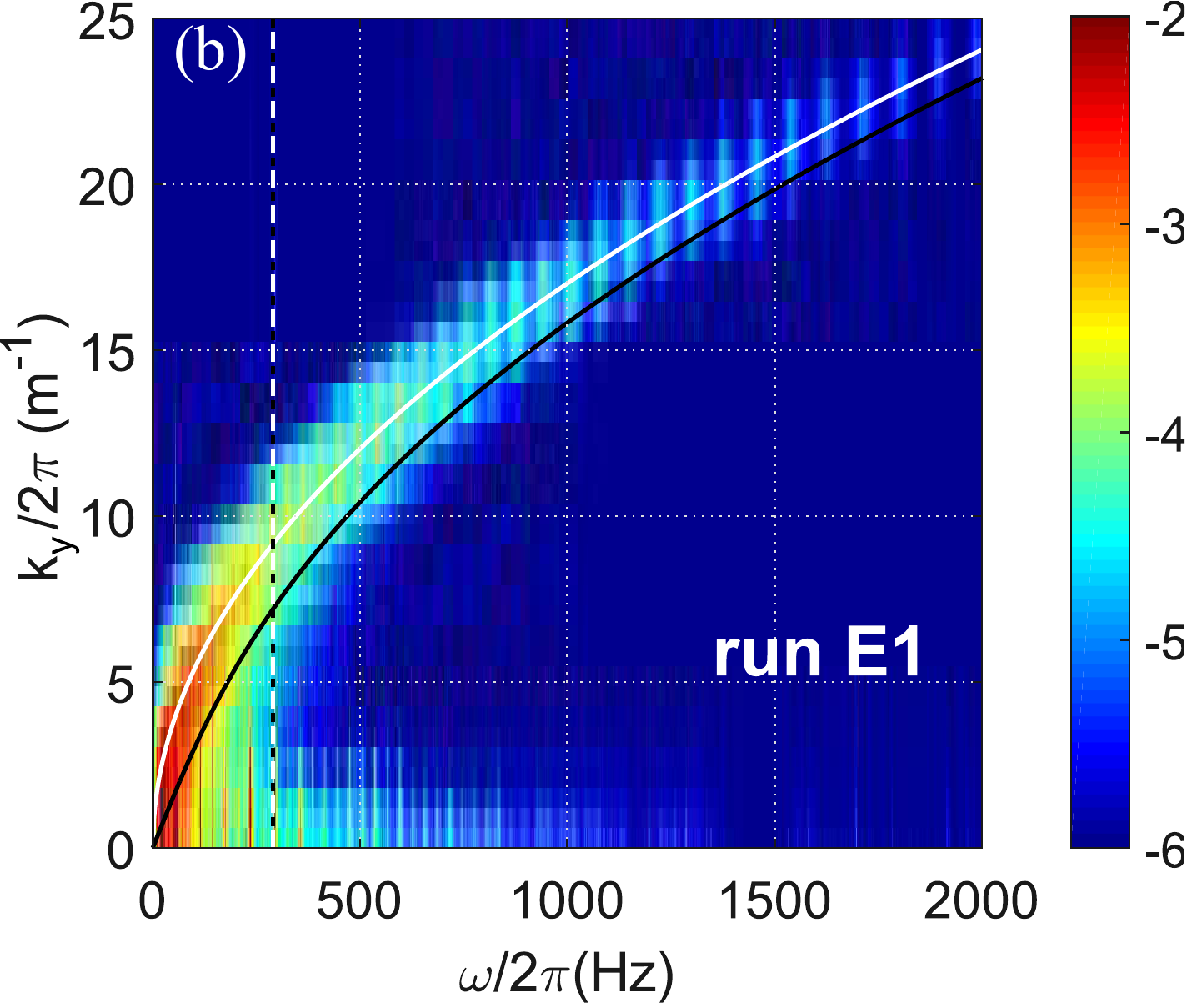}

\includegraphics[clip,width=8cm]{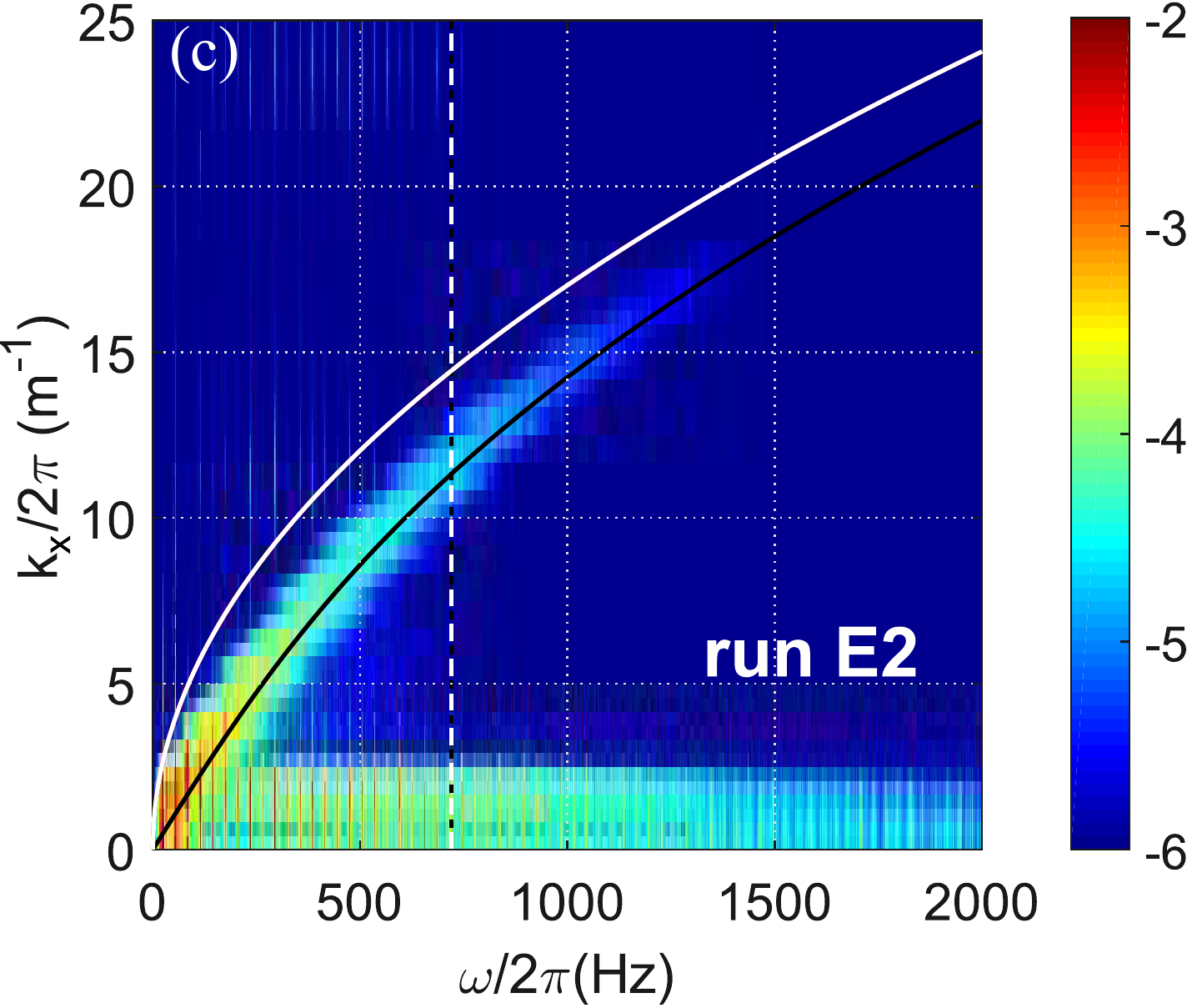}\hfill
 \includegraphics[clip,width=8cm]{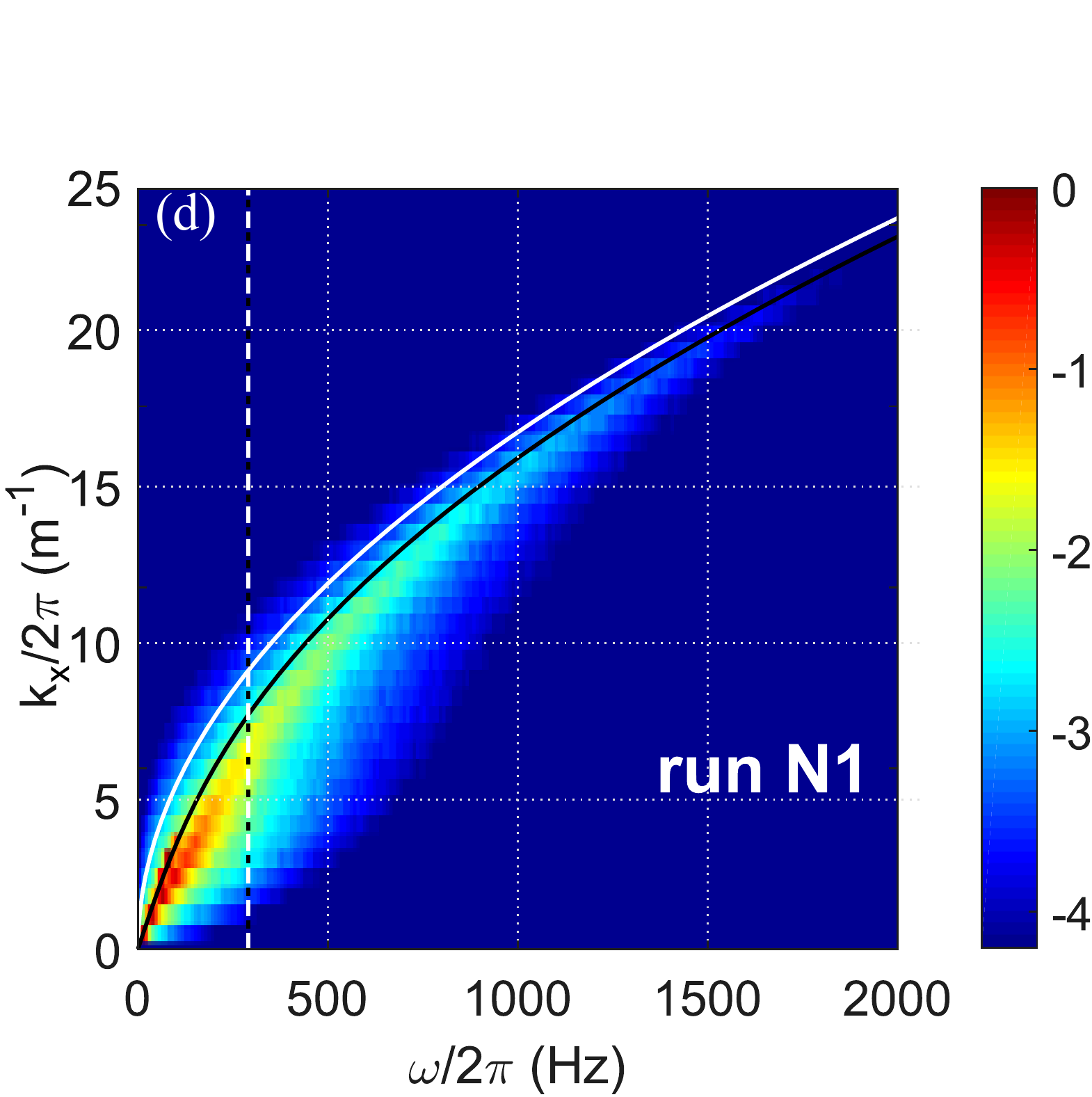}
\caption{(a) normalized spatio-temporal spectrum $E^v(k_x,0,\omega)/E^v(k_{f},0,\omega_f)$ of the velocity field in the stretching direction for experiment E1 with $T_x=4$~kN/m and $T_y=0$. The crossover frequency $\omega_c/2\pi=290$~Hz is shown as a vertical dashed line. (b) same dataset but the spectrum $E^v(0,k_y,\omega)/E^v(0,k_{f},0,\omega_f)$ is shown.
(c) $E^v(k_x,0,\omega)/E^v(k_{f},0,\omega_f)$ for experiment E2 with a larger value of the tension $T_x=10$~kN/m and the maximum amplitude forcing permitted by the electromagnetic shaker. The crossover frequency $\omega_c/2\pi=725$~Hz is shown as a vertical dashed line. (d) $E^v(k_x,0,\omega)/E^v(k_{f},0,\omega_f)$ for a numerical simulation N1 with $T_x=4$~kN/m and $C^{2}_{exp}$ indented to reproduce qualitatively the experiment E1. In all cases the black line corresponds to the dispersion relation taking into account the tension (\ref{dispersion.tension}) and the red line corresponds to the dispersion relation without tension ({rdnotens}).}
\label{spectre_st_exp_num}
\end{figure*}

In this part, experiments were carried out with two values of the applied tension $T_x=4$~kN/m (E1) and $T_x=10$~kN/m (E2). These values of $T$ correspond to a crossover frequency $\omega_c/2\pi$ equal to 290~Hz and 725~Hz respectively. A numerical simulation (N1) was also performed using parameters close to the experiment E1 so that to permit a qualitative comparison between the two. 

In order to study the energy distribution in the media, a Fourier analysis is applied to the deformation and normal velocity ($v=\frac{\partial \zeta}{\partial t}$) of the waves in the $x-y$ space and in time. This yields a 
frequency-wavenumber spectra of the normal velocity noted $E^v(k_x,k_y,\omega)$. We use the spectrum of the normal velocity rather than the one of the deformation $E^\zeta$ for convenience. Indeed the decay of $E^v(k_x,k_y,\omega)$ with the frequency or wavenumber is slower than that of $E^\zeta$. The relation between the two spectra is simply $E^v(k_x,k_y,\omega)=\omega^2E^\zeta(k_x,k_y,\omega)$. In the following, we often show the spectrum summed over the direction of the wavevector $\mathbf k$ which is noted $E^v(k,\omega)$.

Figure~\ref{spectre_st_exp_num}(a) shows the spectrum $E^v(k_x,0,\omega)$ of the wave propagating along the direction of stretching for experiment E1. Figure~\ref{spectre_st_exp_num}(b) shows the spectrum in the $y$ direction orthogonal to the stretching, along which no tension is applied. In both cases the energy is localized along a curved line. In (a) the energy is localized on the linear dispersion relation (\ref{dispersion.tension}) that takes into account the tension. By contrast in (b), the energy is localized on the dispersion relation without tension (\ref{rdnotens}). In fig.~\ref{spectre_st_exp_num}(c), we show the experiment E2 with a value of the tension which is $2.5$ times higher. Due to the increased tension, the energy falls on a dispersion relation that is more shifted from the one with $T=0$. No sign of a qualitative change of behavior can be seen. For instance in~\cite{hassaini2017transition} the change from weak turbulence to soliton was visible in the fact that the energy was lying on a straight line over the whole range of frequencies. Here the energy remains on the predicted dispersion relation which is now anisotropic due to the stretching which is applied only in the $x$ direction. The energy cascades to higher frequencies in a way consistent with the phenomenology of the WTT. For E2 (highest tension), the cascade seems to stop at lower frequencies. Due to the strong stretching the shaker cannot operate at large amplitude and it may affect the injection. It is also possible that dissipation is altered as well.

The numerical simulation N1 shown in fig.~\ref{spectre_st_exp_num}(d) is also qualitatively consistent with the WTT picture. Note that the energy lies also on a line that is actually even below the linear dispersion relation with tension for intermediate values of the frequency (around 400~Hz). This extra shift is actually predicted by the WTT and was also observed in plates without tension~\cite{epjb}. In the simulation, the non linearity must be stronger than in the experiment, possibly due to distinct forcing and boundary conditions. 

The experimental and numerical results shown in this part do not show hints of any qualitative change of dynamics of the vibrations of the plate. The spectra remain qualitatively consistent with what is expected from the WTT. No hint of localized structures such as shocks or solitons can be observed. In comparison with the unstretched case, the only difference is the change of the dispersion relation due to the contribution of the tension in (\ref{dispersion.tensionAnisotropic}). 

In experiments, the only parameter that can be tuned is the strength of stretching $T$. In order to gradually evolve towards the pure membrane, we would rather change the $C^2$ parameter by changing the thickness of the plate. This cannot be achieved experimentally as no plate thinner than $0.5$~mm could be found when keeping the same size. Thus in next part we use numerical simulations to investigate the limit $C^2$ going to zero.

\subsection{Numerical investigation of the transition to the pure membrane: $C^{2}\rightarrow 0$ }

\begin{figure*}[!htb]
\includegraphics[clip,width=8cm]{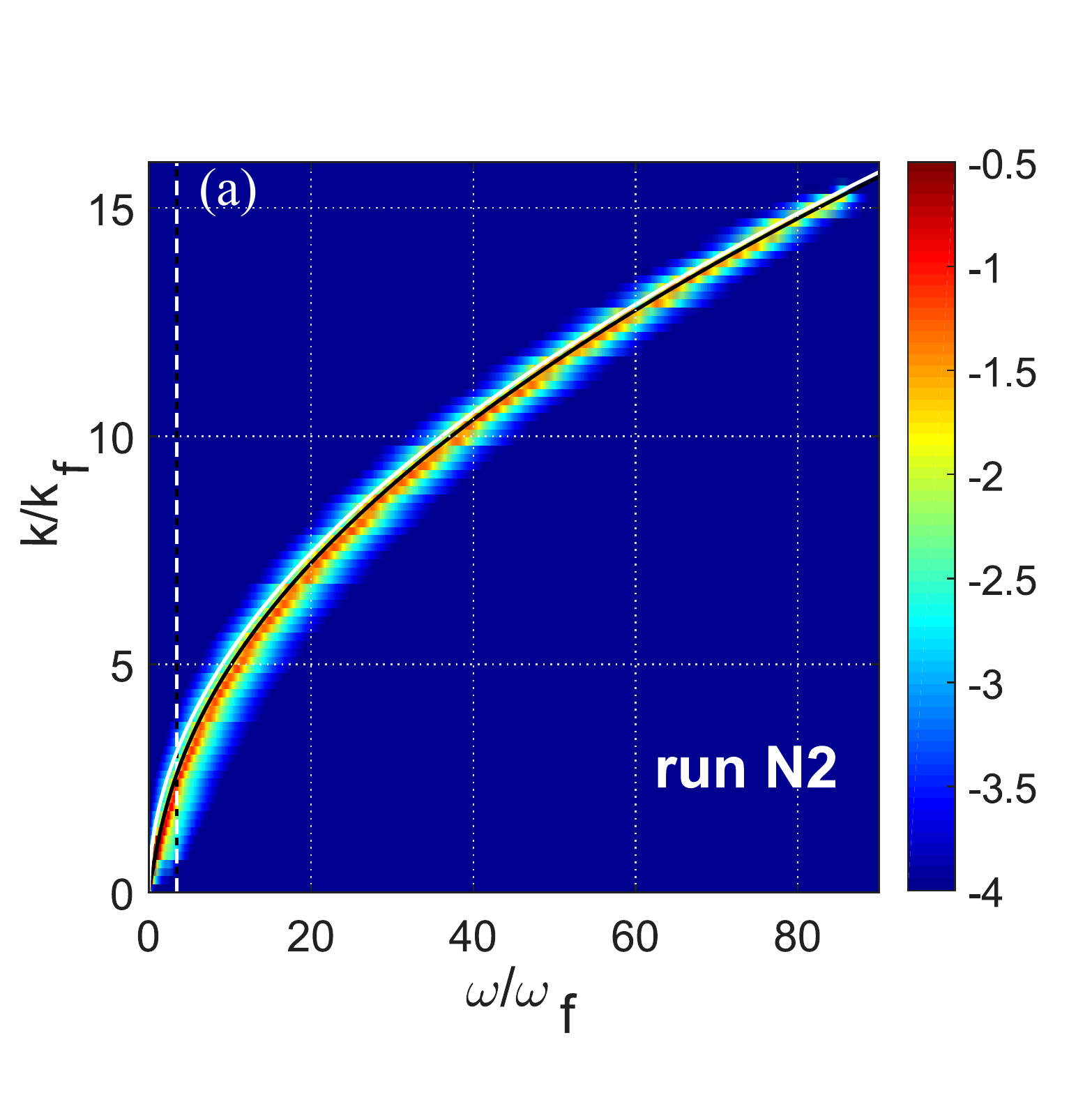}\hfill
\includegraphics[clip,width=8cm]{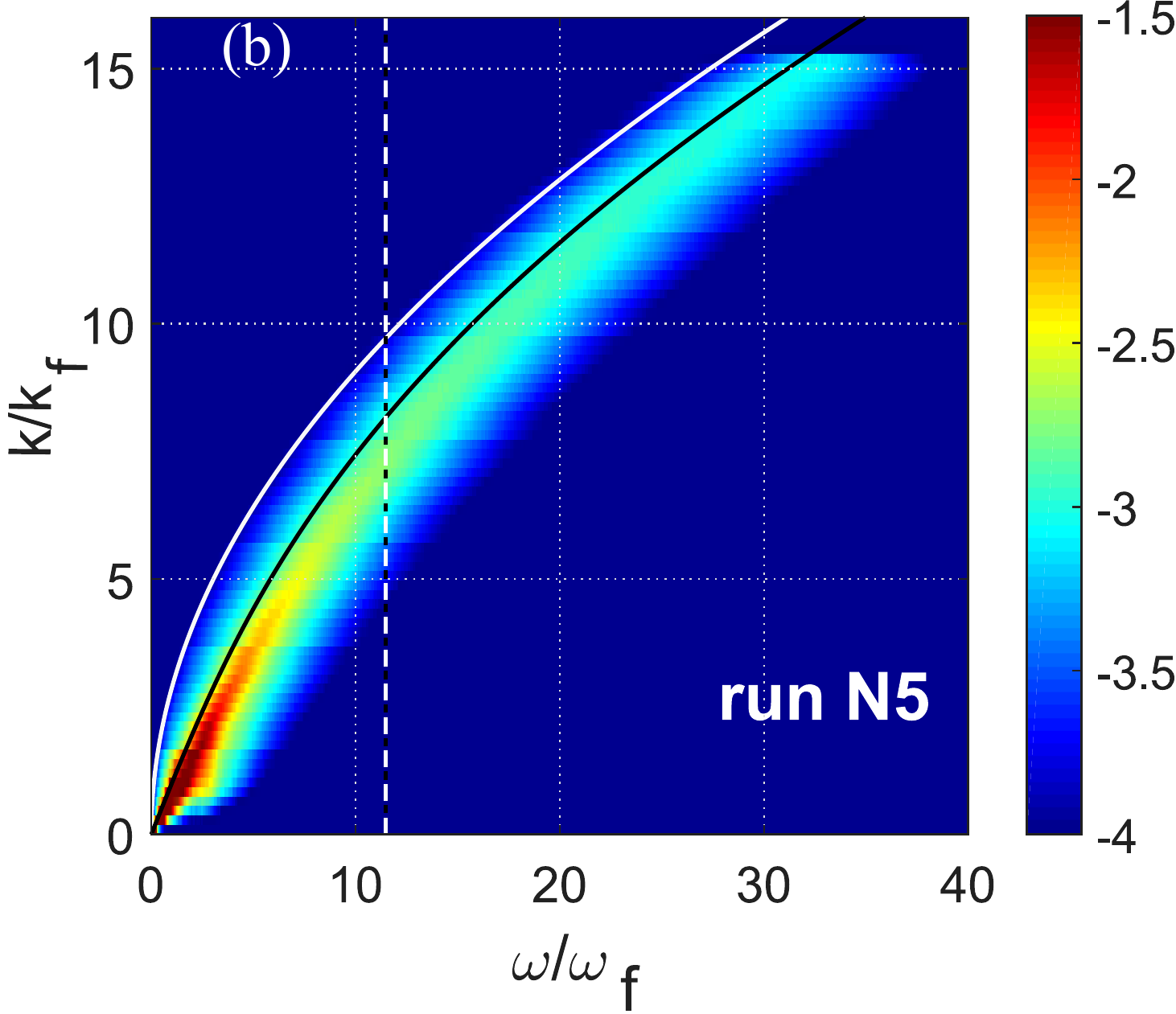}

\includegraphics[clip,width=8cm]{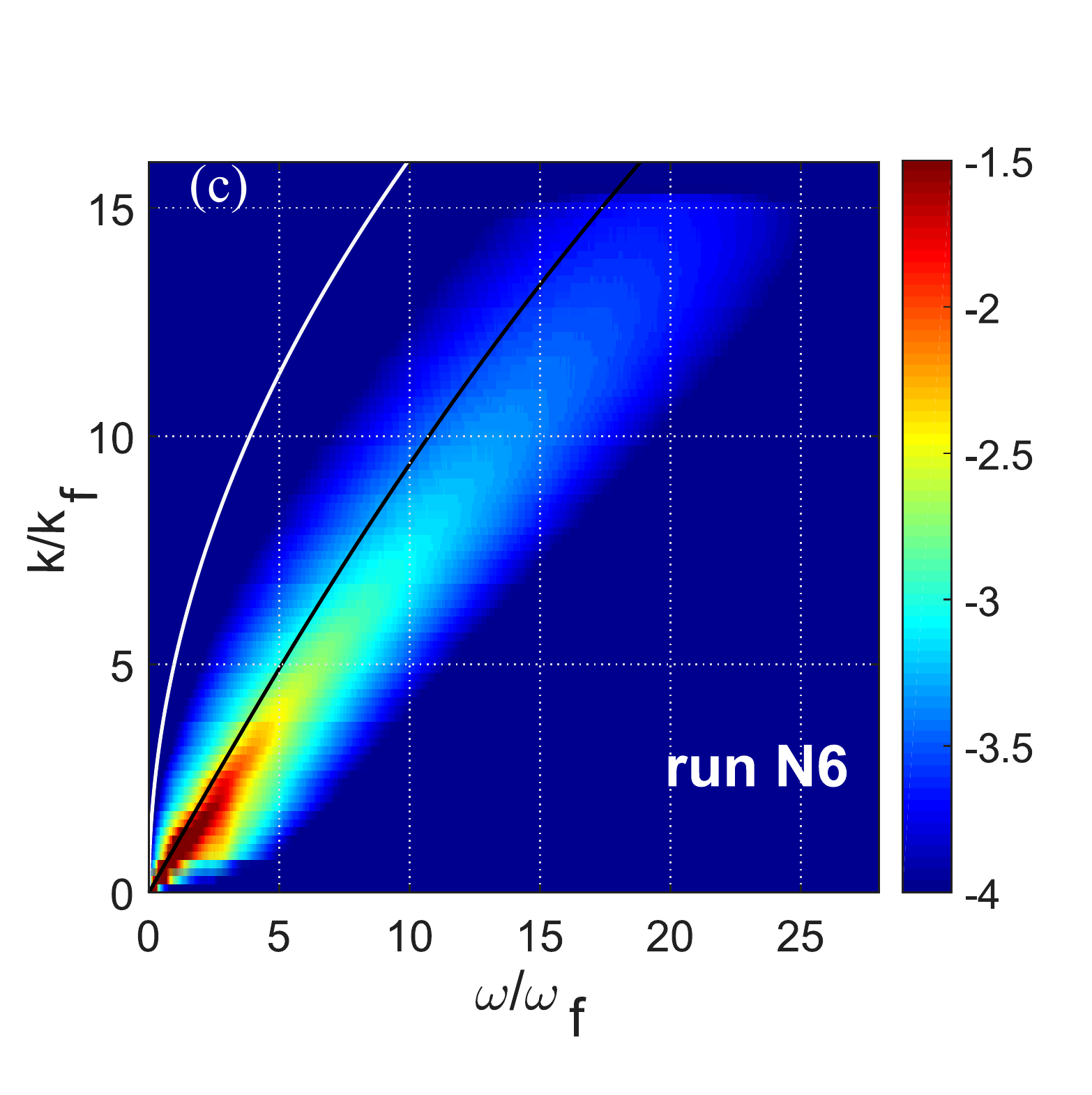}\hfill
\includegraphics[clip,width=8cm]{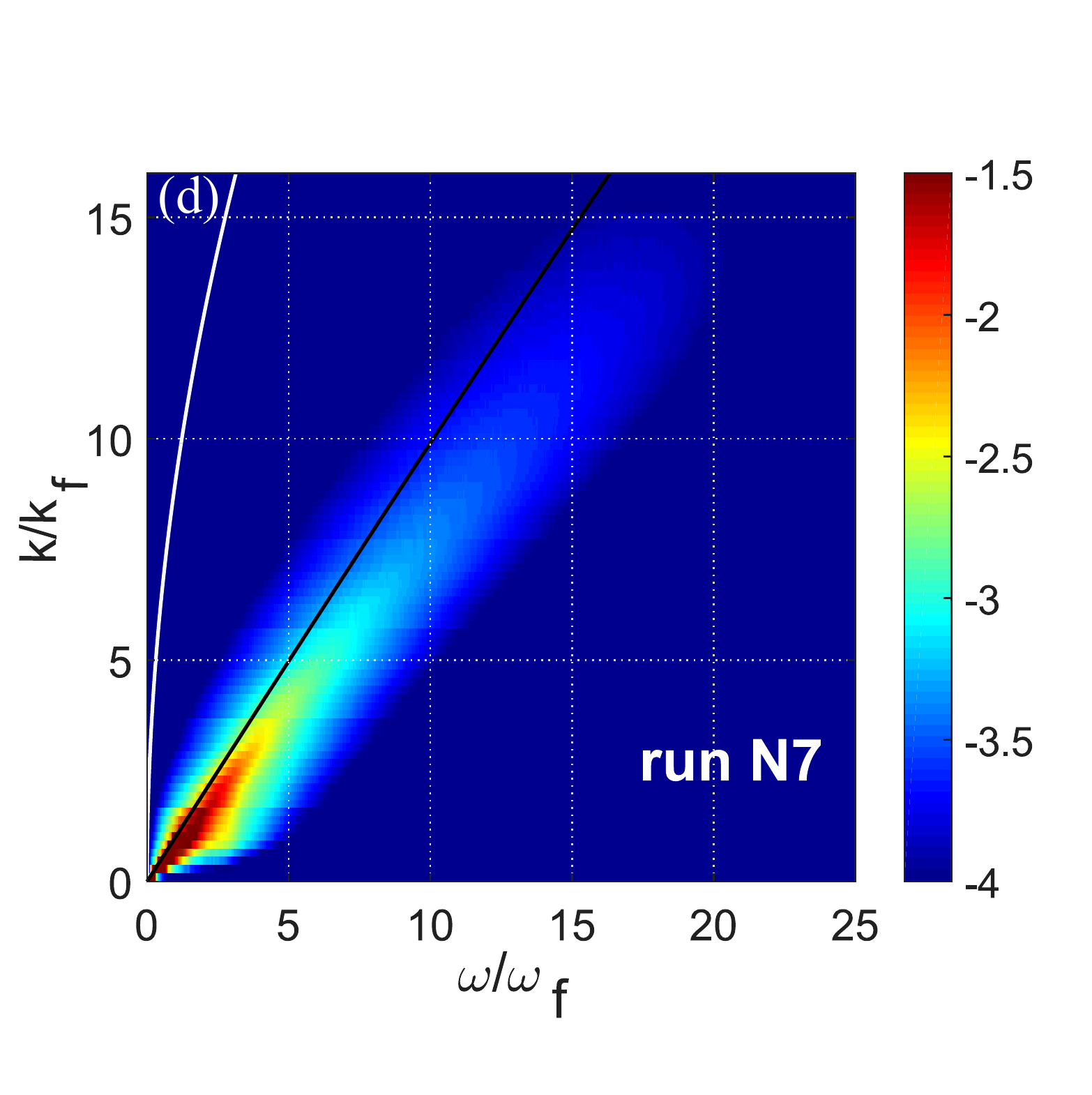}
\caption{Normalized spatio-temporal spectrum of the velocity field $E^v(k,\omega)/E^v(k_{f},\omega_f)$ for the DNS: (a) N2 with $C^{2}_{exp}$ (b) N5, $C^{2}_{exp}/10$ (c) N6, $C^{2}_{exp}/100$ ans (d) N7, $C^{2}_{exp}/1000$. In all cases the black line corresponds to the dispersion relation (\ref{dispersion.tension}) and the red line corresponds to the dispersion relation without tension (\ref{rdnotens}). In (a) and (b) the vertical dashed line shows the corssover frequency $\omega_c$ (which is not in the displayed range for (c) and (d)).}
\label{spectre_st_c2}
\end{figure*}

We examine the effects of dispersion on wave turbulence by running numerical simulations with a constant tension and decreasing gradually the value of $C^{2}$ from the experimental value $C^{2}_{exp}=0.6084$ down to $C^{2}={C^{2}_{exp}}/{1000}$ (N2 to N7). The wavenumber of the forcing is kept constant so that in the range of resolved wavenumbers, the waves are less and less dispersive when $C^2$ is reduced. The amplitude of the forcing is kept constant  and the resulting {\it rms} steepness of the plate is also almost constant close to the value $2.3 \pm0.3\%$. The tension is now applied isotropically in the $x$ and in the $y$ direction and we use the ``ideal dissipation" scheme in order to remove the impact of wideband dissipation in the inertial range and to be in the theoretical framework.

We compare the spatio-temporal spectra $E^v(k,\omega)$ for the case of the experimental $C^{2}_{exp}$ down to $C_{exp}^{2}/1000$ in fig.~\ref{spectre_st_c2} (for a given value of the isotropic tension $T$). We observe that in all cases, the energy is concentrated around the dispersion relation taking into account the tension such as stated in (\ref{dispersion.tension}). This means that even though the media is being less and less dispersive, for all the values of $C^{2}$ tested, the vibration of the plates remains in a state of weak turbulence.

\begin{figure*}[!htb]
\includegraphics[clip,width=8cm]{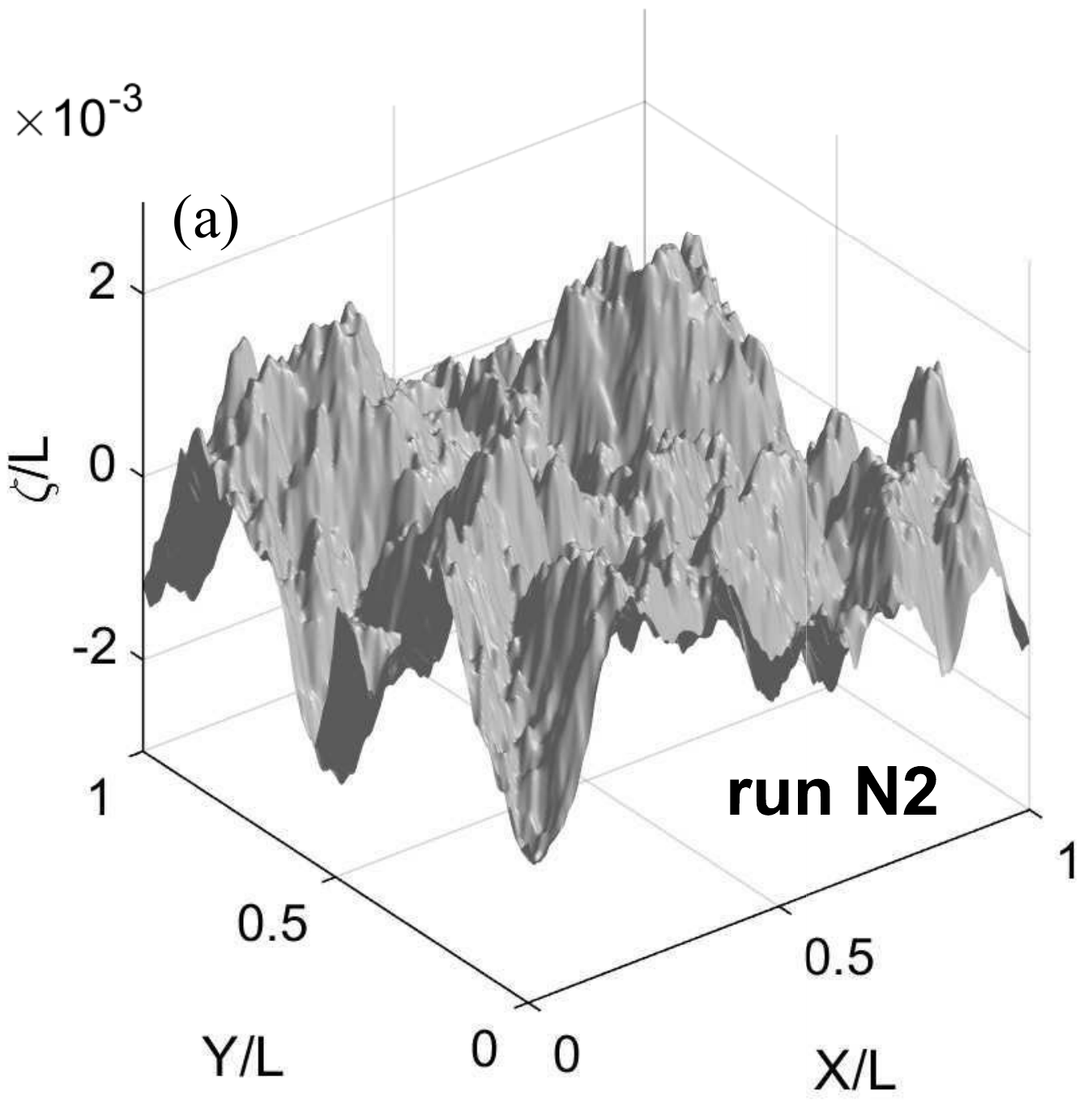}\hfill
\includegraphics[clip,width=8cm]{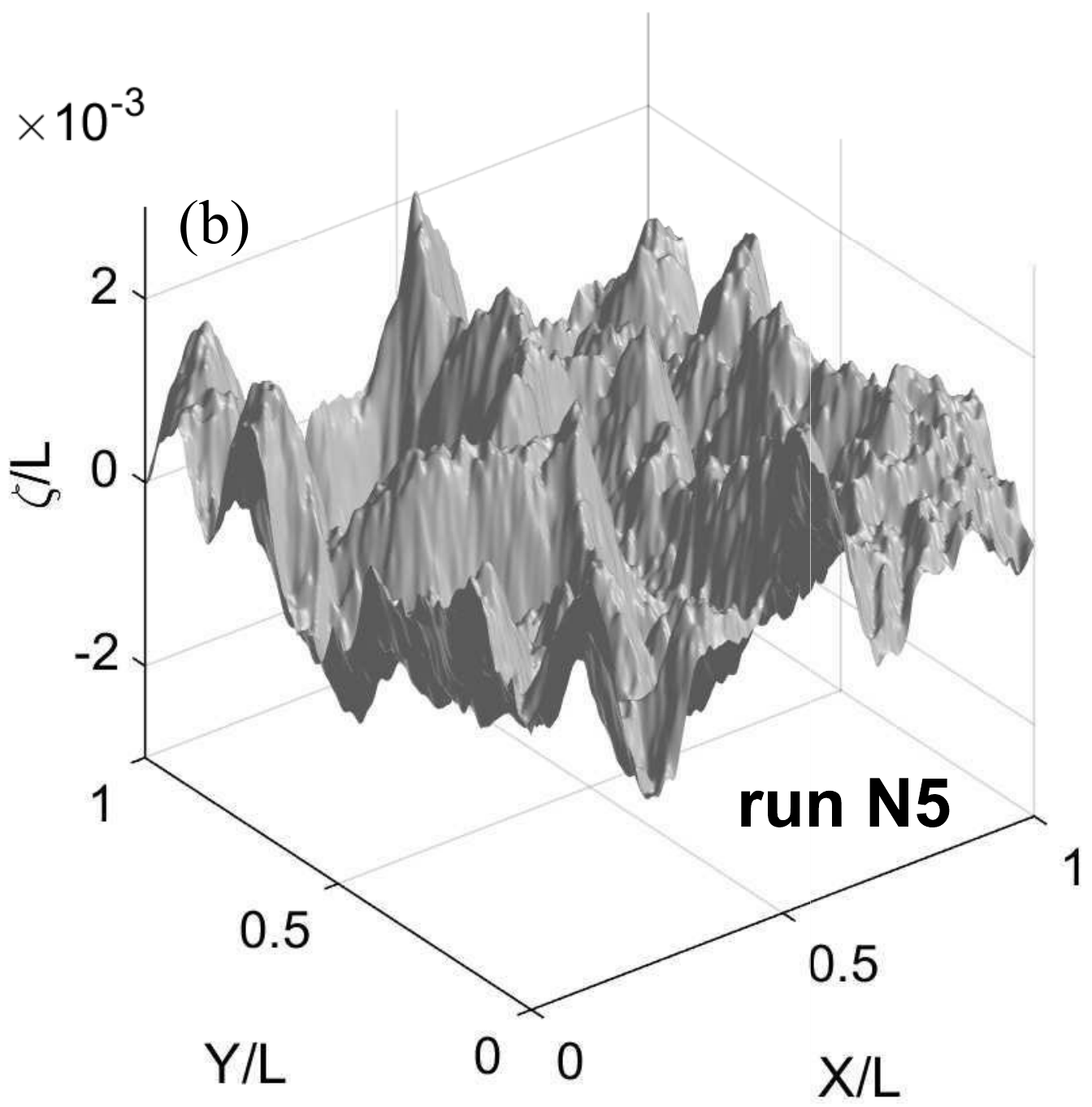}

\includegraphics[clip,width=8cm]{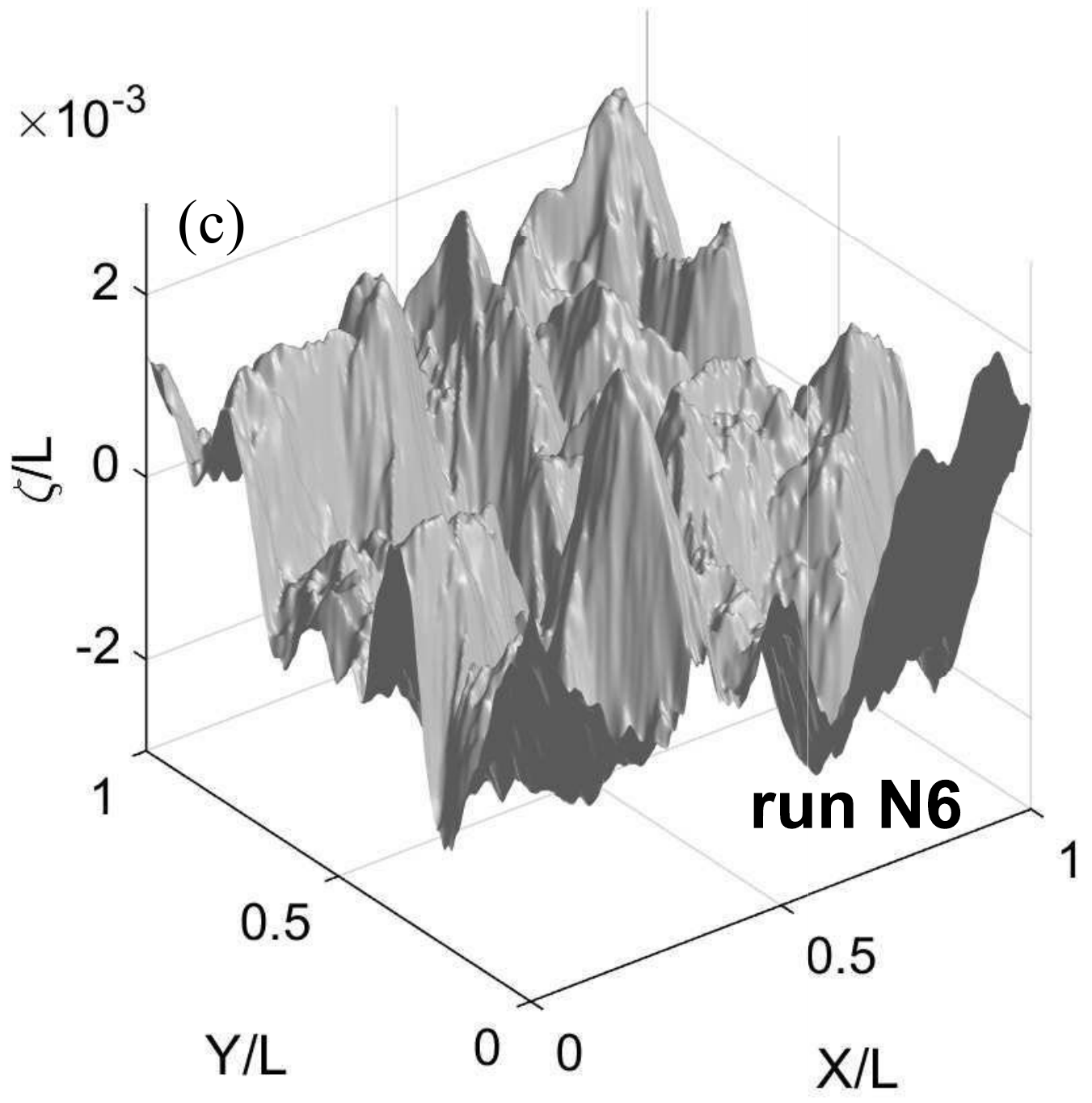}\hfill
\includegraphics[clip,width=8cm]{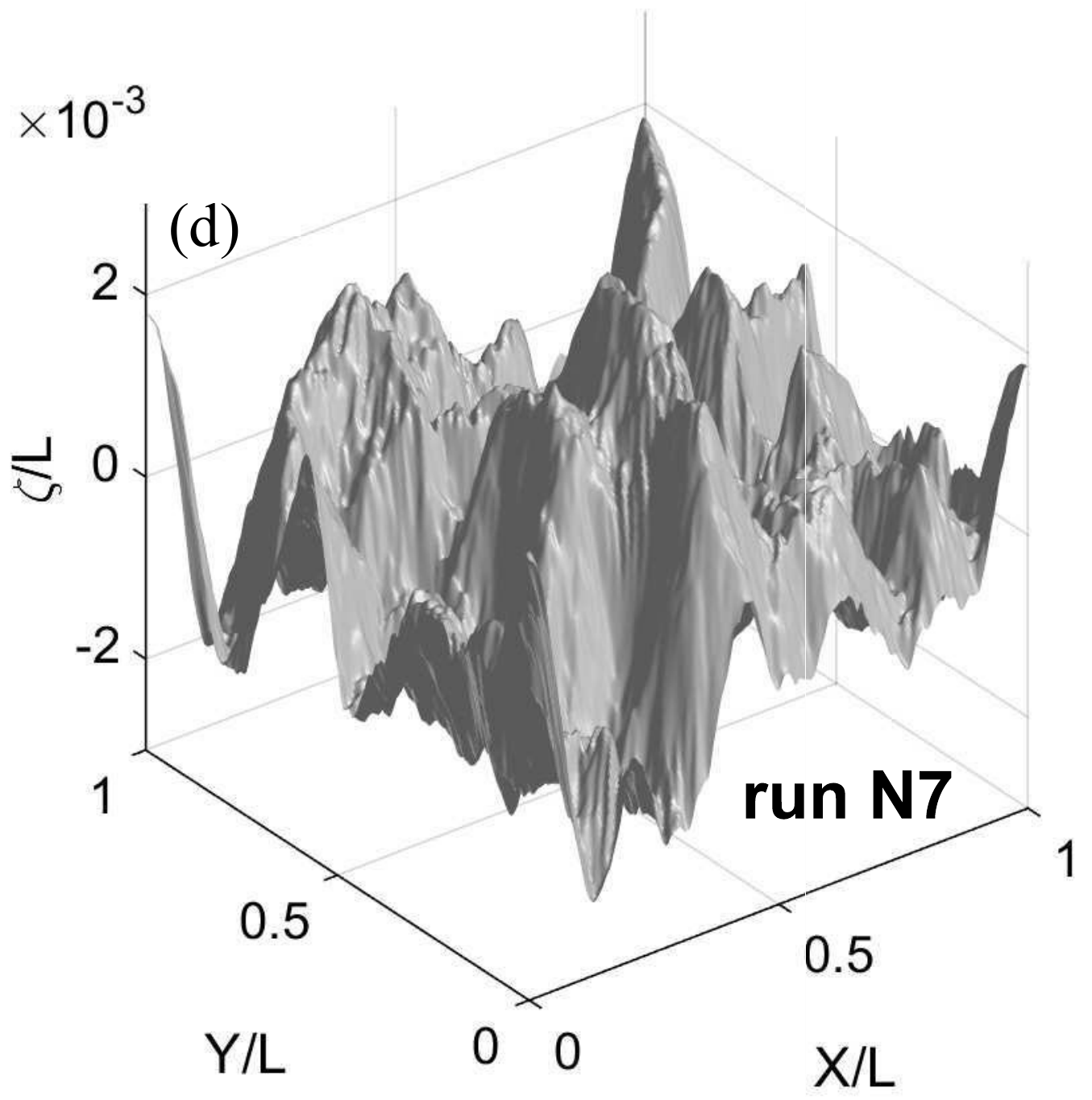}
\caption{Snapshots of the deformation of the plate for the DNS: (a) N2 with $C^{2}_{exp}$ (b) N5, $C^{2}_{exp}/10$ (c) N6, $C^{2}_{exp}/100$ ans (d) N7, $C^{2}_{exp}/1000$. }
\label{surf_weak}
\end{figure*}
 
Snapshots of the surface elevation of the same numerical simulations shown in fig.~\ref{spectre_st_c2} 
 are represented in fig.~\ref{surf_weak}. At all values of $C^2$, the snapshots show a random wave distribution and no extreme events seem to appear.

\begin{figure}[!htb]
\centering
 \includegraphics[width=9.5cm]{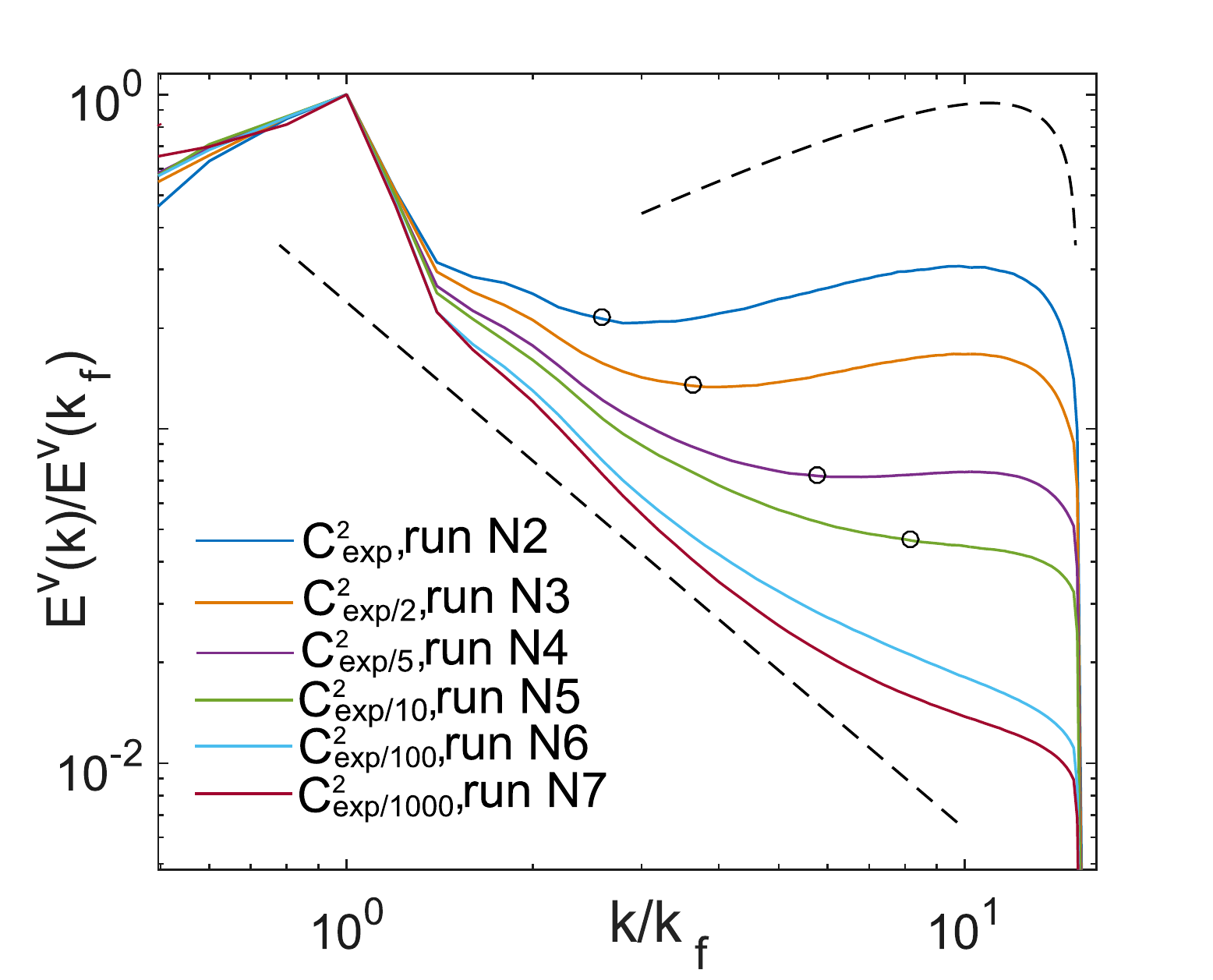}
\caption{$E^v(k)/E^v(k_f)$ for DNS N2 to N7 i.e. for $C^{2}$ decreasing from $C^{2}_{exp}$ to $C^{2}_{exp}/1000$ with a constant tension. The slope of the deformation is kept roughly constant equal to about 2\%. The circles correspond to the crossover wavenumber $k_c$ for the corresponding DNS (when it is small than the dissipative cutoff $k_d=15k_f$). The bottom dashed line shows the theoretical scaling predicted by WTT applied to the pure membrane ($C^2=0$) $E^v(k)\propto k^{-4/3}$. The top dashed line shows the theoretical shape $k\log^{1/2}(k_d/k)$ expected from the WTT applied to the plate~\cite{during}.}
\label{E_k_v}
\end{figure}
In order to have a quantitative information of the decay of the spectrum with the wavenumber, we extract the wavenumber-only spectrum of the normal velocity $E^v(k)$ for the DNS N2 to N7 (fig.~\ref{E_k_v}). The spectrum is seen to steepen when $C^2$ is reduced. At the highest value of $C^2$ comparable to the one of the experimental steel plate, the spectrum is consistent at high wavenumber with the prediction of the WTT for a plate without tension. When decreasing $C^2$, the shape of the spectrum evolves by developing a steeper region at $k$ below $k_c$. In the three larger values of $C^2$ for which $k_c$ remains moderate, the spectrum at $k>k_c$ is qualitatively consistent with the prediction for the non-stretched plate.  Indeed at $k>k_c$ the correction to the no-tension dispersion relation is negligible and this explains why the spectrum keeps a shape compatible with the theory without tension. At the lowest value of $C^2$ (lowest curve), the spectrum is significantly steeper compatible with a $k^{-4/3}$ decay at intermediate values of $k$. This scaling is actually consistent with the WTT prediction of the pure membrane with $C^2=0$ (see \ref{DCDNS} below). At high $k$ a departure from the $k^{-4/3}$ scaling is observed which is probably a reminiscent effect of dispersion as $k_c$ remains finite (although larger than the dissipative cutoff). 

Thus we have experimental and numerical support of the fact that a pure membrane with $C^2=0$ should still exhibit a regime of weak turbulence. These observations must be contrasted with the case of surface water waves for which a transition to a solitonic regime was observed when reducing the dispersion of the waves (by reducing the water depth)~\cite{hassaini2017transition}. In the next section, we focus on the asymptotic case of the pure membrane for which $C^2=0$. We first apply the WTT formalism to this case. Then we compare its predictions with numerical simulations.

\section{The vibrating non linear elastic membrane}
\label{pure}

\subsection{Application of the WTT to the membrane}

\subsubsection{Kinetic equation}

From the modifed F\"oppl-von K\'arm\'an equations (\ref{foppl1}) and (\ref{fopplTension}), one notices that only the linear term differs from the well studied case of elastic plates without tension. Furthermore, one can show that the so-called kinetic equation (equation for the slow temporal evolution of the wave spectrum) obtained for elastic plates under tension only differs in the dispersion relation with the one obtained for $T=0$ in \cite{during}. We shall consider the Fourier transform $$\zeta({\bf r},t)=\frac{1}{2\pi}\int \zeta_{\bf k}e^{i{\bf k}\cdot{\bf r}}d{\bf k},$$ 
and the canonical transformation 
\begin{equation}
\zeta_{\bf k}=\frac{X_k}{\sqrt{2}}(A_{\bf k}+A^*_{-\bf k})
\label{CT}
\end{equation}
 with $X_k={1}/{\sqrt{\omega_k\rho h}}$. In the weakly non-linear limit, a kinetic equation for the evolution of the wave action spectrum $n_{\bf k}=\langle A_{\bf k} A^*_{\bf k}\rangle/L^2$ can be obtained. The mean value $\langle \,\cdot\,\rangle $ represents an ensemble average and $L$ is the size of the plate. 
Formally,  the kinetic equation is given by (for a detailed derivation see \cite{platesPhysD}) 
\begin{multline}
\frac{dn_{\bf p}}{dt}=12\pi \sum_{s_1s_2s_3=\pm1} \int |J_{-{\bf p} {\bf k}_1{\bf k}_2{\bf k}_3} |^2  n_{{\bf k}_1}n_{{\bf k}_2}n_{{\bf k}_3}n_{{\bf p}}\\
\times\left(\frac{1}{n_{\bf p}}-\frac{s_1}{n_{{\bf k}_1}}-\frac{s_2}{n_{{\bf k}_2}}-\frac{s_3}{n_{{\bf k}_3}}\right)\delta({\bf k}_1+{\bf k}_2+{\bf k}_3-{\bf p})\\ 
\times \delta(s_1\omega_{{\bf k}_1}+s_2\omega_{{\bf k}_2}+s_3\omega_{{\bf k}_3}-\omega_{\bf p}) d{\bf{k}}_{123},
\label{kinetic1}
\end{multline}
where the scattering matrix introduced in \cite{during} is given by
\begin{multline}
J_{{\bf p} {\bf k}_1{\bf k}_2{\bf k}_3} = \\\left(\frac{1}{|{\bf p}+{\bf k}_1|^4}+\frac{1}{|{\bf k}_2+{\bf k}_3|^4}\right)\frac{|{\bf p}\times {\bf k}_1|^2 |{\bf k}_2\times {\bf k}_3|^2}{48 \rho^2h^2\sqrt{\omega_{\bf p}\omega_{{\bf k}_1}\omega_{{\bf k}_2}\omega_{{\bf k}_3} }}\\+{\rm permutations\,of\,indices\,}1,2,3.\label{Eq:ScatMat}
\end{multline}

In order to derive the kinetic equation above, the existence of a resonant manifold is assumed. Namely, a manifold in wave-vector space that satisfies
\begin{equation}
h({\bf k}_1,{\bf k}_2,{\bf p})=s_1\omega({\bf k}_1)+s_2\omega({\bf k}_2)+s_3\omega({\bf k}_3)-\omega({\bf p})=0,
\end{equation}
with $s_1{\bf k}_1+s_2{\bf k}_2+s_3{\bf k}_3+{\bf p}=0$. In general, an extra technical condition needed to perform a saddle-point like approximation is required. It is assumed that gradients of $h({\bf k}_1,{\bf k}_2,{\bf p})$ have to be different from zero on the resonant manifold. Under these assumptions, the multi-scale asymptotic expansion used to derive the kinetic equation remains bounded  \cite{Saffman, platesPhysD} (see Appendix \ref{Saddle-node}). The kinetic equation contains two different groups of terms depending on the type of resonance: the $2\leftrightarrow 2$ interaction (eg. $s_1=-1$ and $s_2=s_3=1$) and the $3\leftrightarrow 1$ interaction (eg. $s_1=s_2=s_3=1$).  The $2\leftrightarrow 2$ interaction is common in systems that are phase invariant (e.g Non Linear Schr\"odinger equation) or have a non-decay dispersion relation ($\omega_k\sim k^\alpha$ with $\alpha<1$), such systems have an extra conserved quantity named the wave action $N=\int n_{\bf p}{\mathrm d}{\bf p}$. The $3\leftrightarrow 1$ type of interaction is less common, since require a decay dispersion relation $\alpha\geq1$ and the symmetry of the Hamiltonian under reflection $\zeta \rightarrow -\zeta$. Such interaction prevents the wave action conservation, thus {\it a priori} wave action is not conserved for the membrane.

As it was first shown by Zakharov, stationary out-of-equilibrium solutions of the kinetic equation can be found~\cite{R1}. Whenever the dispersion relation and the scattering matrix are homogeneous functions, such solutions, known as Kolmogorov-Zakharov spectra, can be obtained analytically. The limit $T\to0$ (no tension)  corresponds to the case of elastic plates, for which the kinetic equation with $\omega_{\bf k} = C  k^2 $ was derived in \cite{during}. In the other limit,  $h^3E\ll T$,  that corresponds to a non-linear membrane, the linear term becomes non dispersive with $\omega_{\bf k} =   \sqrt{ \frac{T}{\rho h}} k$. Special care needs to be taken in this limit as the gradient of $h({\bf k}_1,{\bf k}_2,{\bf p})$ can identically vanish on the resonant manifold for some type of interactions (see Appendix \ref{non-dispersive}). Such interactions seem to be the responsible for the growth of strong non-linearities  that can lead to the formation of shock waves and the breakdown of the wave turbulence theory \cite{aucoin,acustictur}. It remains as an open question under which circumstance non dispersive non-linear wave systems either focalize energy into rays leading to shock wave formation, or redistribute angularly the energy through nonlinearity so that shock wave formation is suppressed.
 
 In general, a non dispersive system with a cubic non-linearity contains $2\leftrightarrow 2$ and $3\leftrightarrow 1$ wave interactions.
We can show that for $2\leftrightarrow 2$ interactions, the resonant manifold contains collinear train waves (parallel or antiparallel waves) but also non-trivial solutions (see appendix \ref{non-dispersive} and Fig.\ref{cone}). In this case, the multi-scale expansion leading to the kinetic equation remains bounded and its derivation is justified. In the case of $3\leftrightarrow 1$ interactions, the resonant manifold only contains collinear solutions and special care is needed. The behavior of collinear interactions seems to be case dependent and will be addressed elsewhere. In the particular case of an elastic membrane, eventual divergences in the multi-scale expansion are healed by the scattering matrix \eqref{Eq:ScatMat}. Indeed, collinear interactions of waves are completely suppressed by the vector products in $J_{{\bf p} {\bf k}_1{\bf k}_2{\bf k}_3}$. It follows that only $2\leftrightarrow 2$ non-collinear interactions contribute to the kinetic equation (\ref{kinetic1}), what ensures its validity (see Appendix \ref{non-dispersive}). The kinetic equation then simplifies to:
 \begin{multline}
\frac{dn_{\bf p}}{dt}=36\pi  \int |J_{-{\bf p} {\bf k}_1{\bf k}_2{\bf k}_3} |^2  n_{{\bf k}_1}n_{{\bf k}_2}n_{{\bf k}_3}n_{{\bf p}}\\
\times\left(\frac{1}{n_{\bf p}}+\frac{1}{n_{{\bf k}_1}}-\frac{1}{n_{{\bf k}_2}}-\frac{1}{n_{{\bf k}_3}}\right)\delta({\bf k}_1+{\bf k}_2+{\bf k}_3-{\bf p})\\
\times \delta(\omega_{\bf p}+\omega_{{\bf k}_1}-\omega_{{\bf k}_2}-\omega_{{\bf k}_3}) d{\bf{k}}_{123}.
\label{kinetic2}
\end{multline}

An important consequence of the fact that in (\ref{kinetic2}) only $2\leftrightarrow 2$  interactions are present, is the conservation of the wave action $N=\int n_{\bf p}{\mathrm d}{\bf p}$. Associated to this (weak) invariant, a new cascade is expected to emerge. Note that $N$ is not an invariant of the full F\"oppl--von K\'arm\'an equations (\ref{fopplTension}), and such a cascade can be expected only within the validity of the weak wave turbulence theory.

\subsubsection{Stationary out-of-equilibrium spectra}

Considering the standard Zakharov transformation two different out-of-equilibrium solutions can be found~\cite{R1,Nazarenko}. The first one corresponds to a constant energy flux $P$ and leads to a direct cascade. It reads $n(\mathbf k)=C_p P^{1/3}k^{-10/3}$ with $C_p$ a constant that could be in principle calculated numerically from the kinetic equation. In terms of the amplitude deformation spectrum, it becomes $\langle |\zeta_{\mathbf k}|^2 \rangle \sim P^{1/3} k^{-13/3}$. If one consider the sum over the angles, it becomes
\begin{equation}
E^\zeta(k)\sim \langle |\zeta_{\mathbf k}|^2 \rangle k\sim P^{1/3} k^{-10/3}\, .
\end{equation}
The second out of equilibrium solution corresponds to a constant flux of wave action $Q$. It leads to an inverse cascade and it reads $n(\mathbf k)=C_q Q^{1/3}k^{-3}$. $C_q$ is also a constant that can be calculated from the kinetic equation. Summing over the angles, we obtain for the spectrum of $\zeta$
\begin{equation}
E^{\zeta}(k)\sim \langle |\zeta_{\mathbf k}|^2 \rangle k\sim Q^{1/3} k^{-3}\, .
\end{equation}

\subsubsection{Stationary equilibrium solutions}
Let us also remind that another statistical steady state exists that corresponds to thermodynamic equilibrium (with no flux). The exact solution in thermodynamic equilibrium has two asymptotes : one corresponding to energy equipartition and the second one corresponding to wave action equipartition. The expected spectra for both cases are:
\begin{eqnarray}
 E_{k}^{\xi} \propto &k^{-1}\quad&\textrm{for energy equipartition}\\
 E_{k}^{\xi} \propto &k^{-2}\quad&\textrm{for wave action equipartition}.
\label{equi_n_membrane}
\end{eqnarray}

\subsection{Direct energy cascade}

\subsubsection{Results}
\label{DCDNS}

\begin{figure}[htb]
\includegraphics[width=0.45\textwidth]{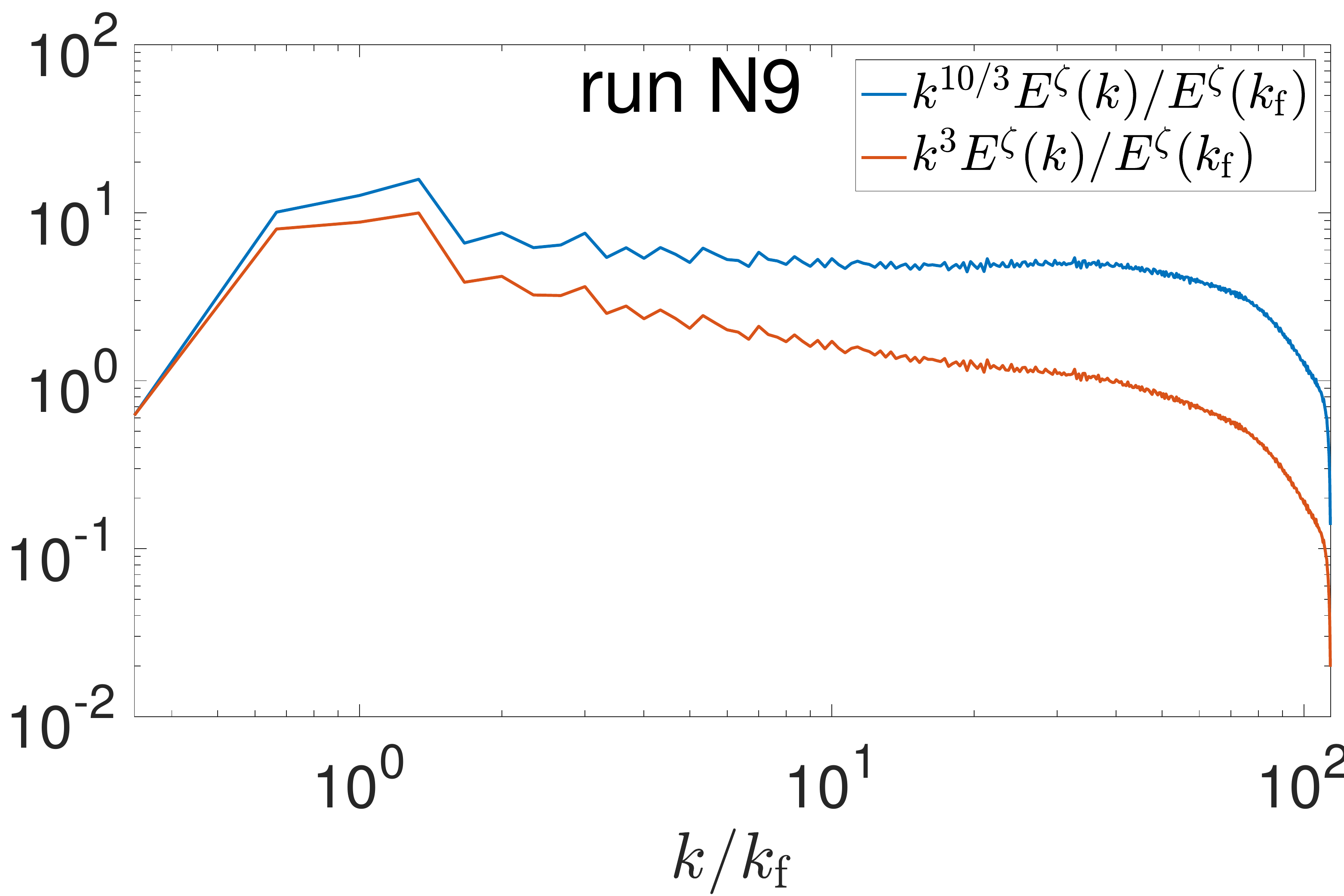}
\caption{Wavenumber spectrum of the deformation $E^{\zeta}(k)$ for the high resolution DNS N9 of the pure membrane. The spectrum has been compensated either by $k^{10/3}$, (which is the WTT prediction) or by $k^3$ (WTT prediction for a plate with $C^2>0$ and $T=0$, discarding the logarithmic corrections). Forcing acts in modes $k\in (1,4)$.}
  \label{k_weak}
\end{figure}
\begin{figure}[htb]
\includegraphics[width=0.45\textwidth]{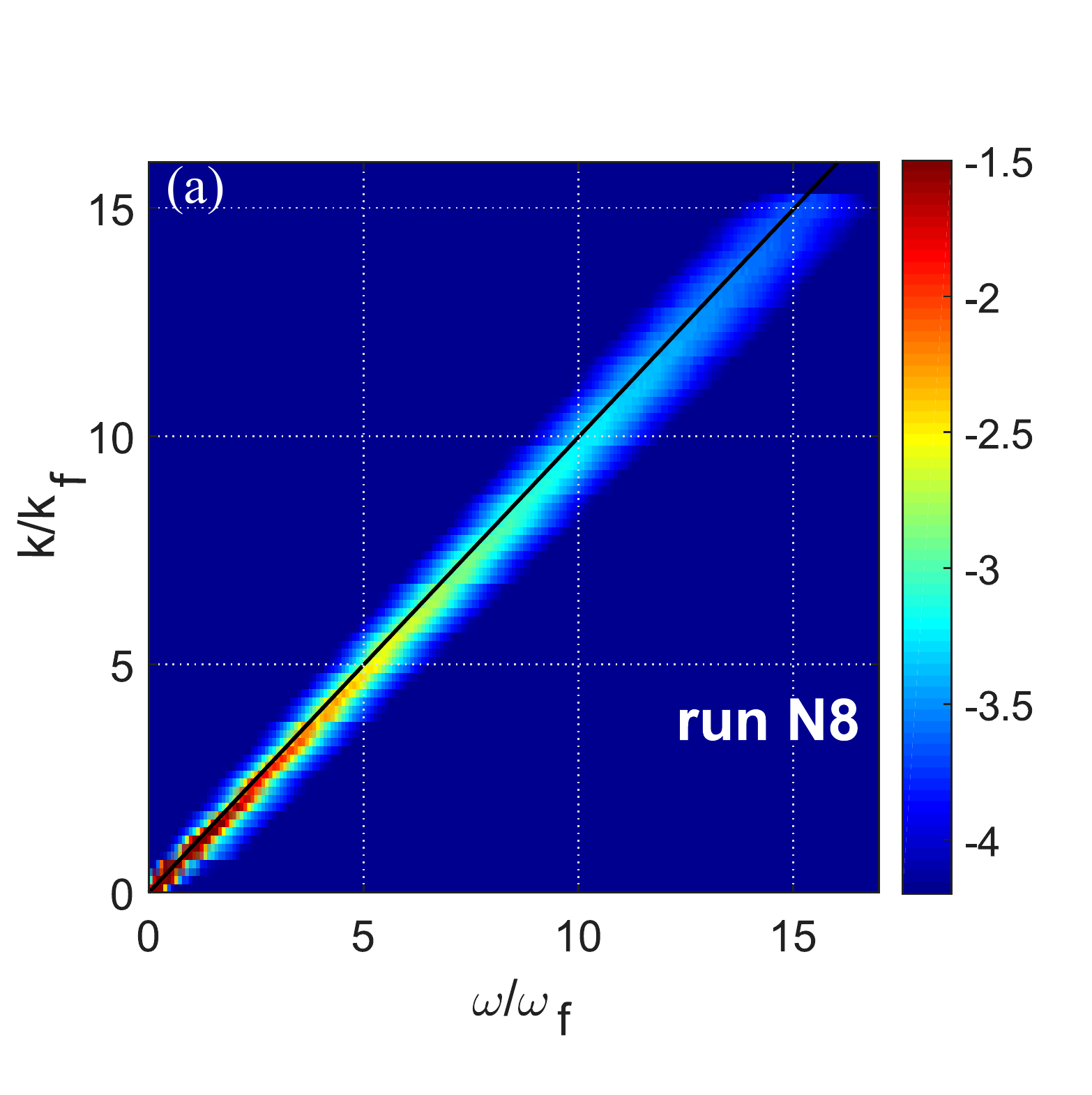} 

\includegraphics[width=8.5cm]{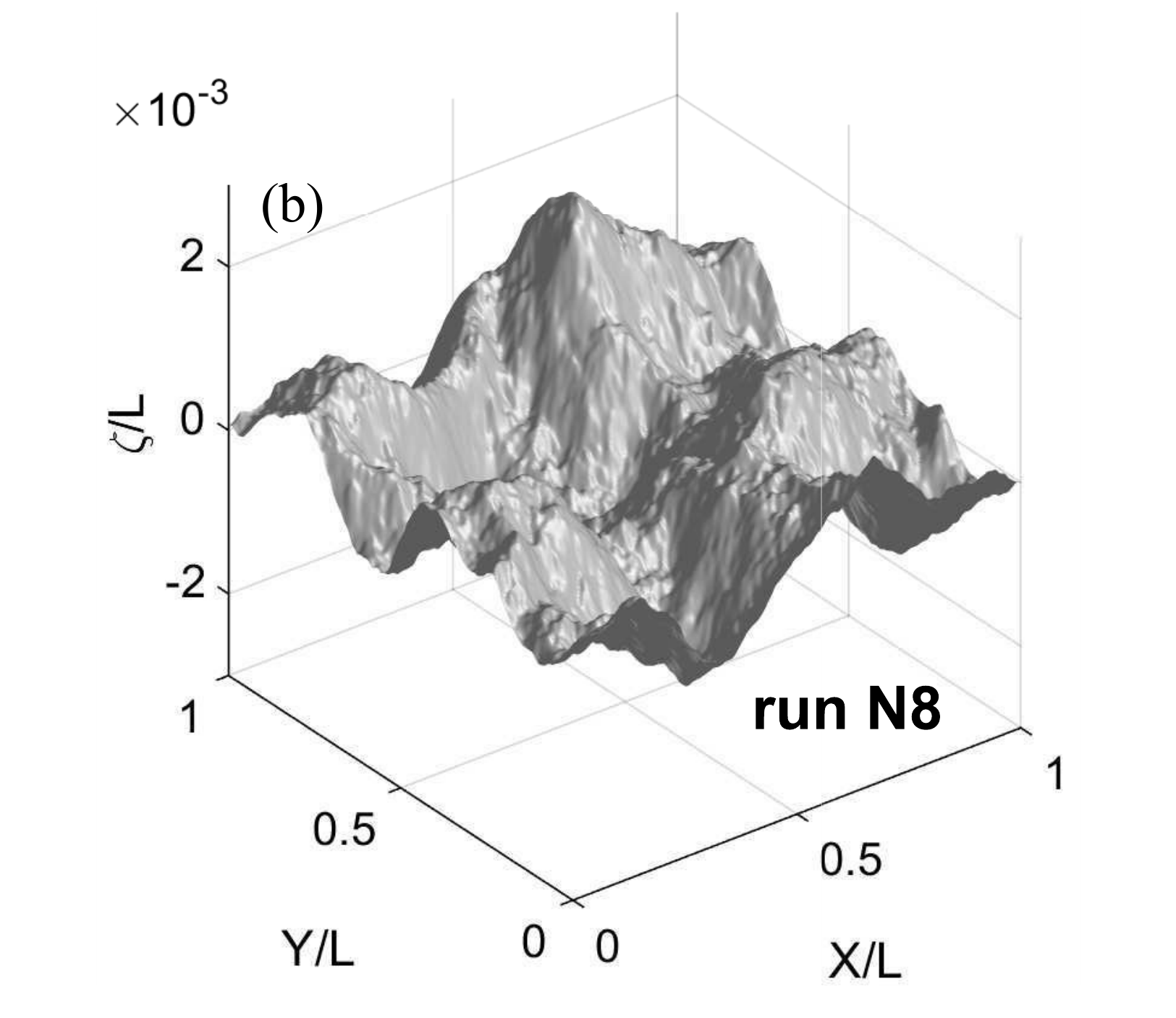}
 \caption{(a) Wavenumber-frequency spectrum $E^{\zeta}$  spectrum of the height of the waves for the same moderate resolution simulation shown in fig.~\ref{k_weak}(a). The black line is the linear dispersion. (b) snapshot of the deformation of the membrane for the same simulation.}
  \label{st_weak}
\end{figure}

As shown in the previous section the WTT prediction for the spectrum of the direct energy cascade is 
$E^{\zeta}(k) \propto k^{-\frac{10}{3}}$ while the case of the tensionless plate is $E^{\zeta}(k) \propto k^{-3}\ln^{1/3}(k^\star/k)$. Discarding the logarithmic corrections, it means that the distinction between the two spectra is quite small. Figure~\ref{k_weak} shows the spectrum for the high resolution DNS N9 of the pure membrane. 
It is consistent with the $-10/3$ spectral exponent expected from the direct cascade in the membrane. The DNS shows unambiguously that the scaling is distinct from the plate one with spectral exponent $-3$. 
Figure~\ref{st_weak}(a) shows the wavenumber-frequency spectrum of the moderate resolution run N11. As expected in weak turbulence, the energy is localized around the linear dispersion relation. The snapshot of the deformation shows that the deformation is totally disordered with no  visible localized singular structures. 
If the spectrum of the deformation is $E^{\zeta}(k) \propto k^{-\frac{10}{3}}$ and the frequency is linear in $k$ then, if the motion is made only of waves, the velocity spectrum should follow: $E^{v}(k)=\omega(k)^2E^{\zeta}(k) \propto k^{-\frac{4}{3}}$. This is the scaling observed in fig.~\ref{E_k_v} at the lowest value of $C^2$.

%
%
%

\subsection{Inverse energy cascade}

In an elastic plate the presence of $3\leftrightarrow1$ wave resonances prevents the conservation of wave action and thus that of a true inverse cascade~\cite{during}. Note that a seemingly inverse cascade was nonetheless reported in numerical simulations~\cite{During:2015gt} but its physical origin remains largely to be explained. For the membrane, as mentioned above, no such $3\leftrightarrow1$ interaction exists so that an inverse cascade related to the conservation of wave action is predicted corresponding to a deformation spectrum $E_{k}^{\xi} \propto k^{-3}$. 

\begin{figure}[!htb]
\includegraphics[width=8cm]{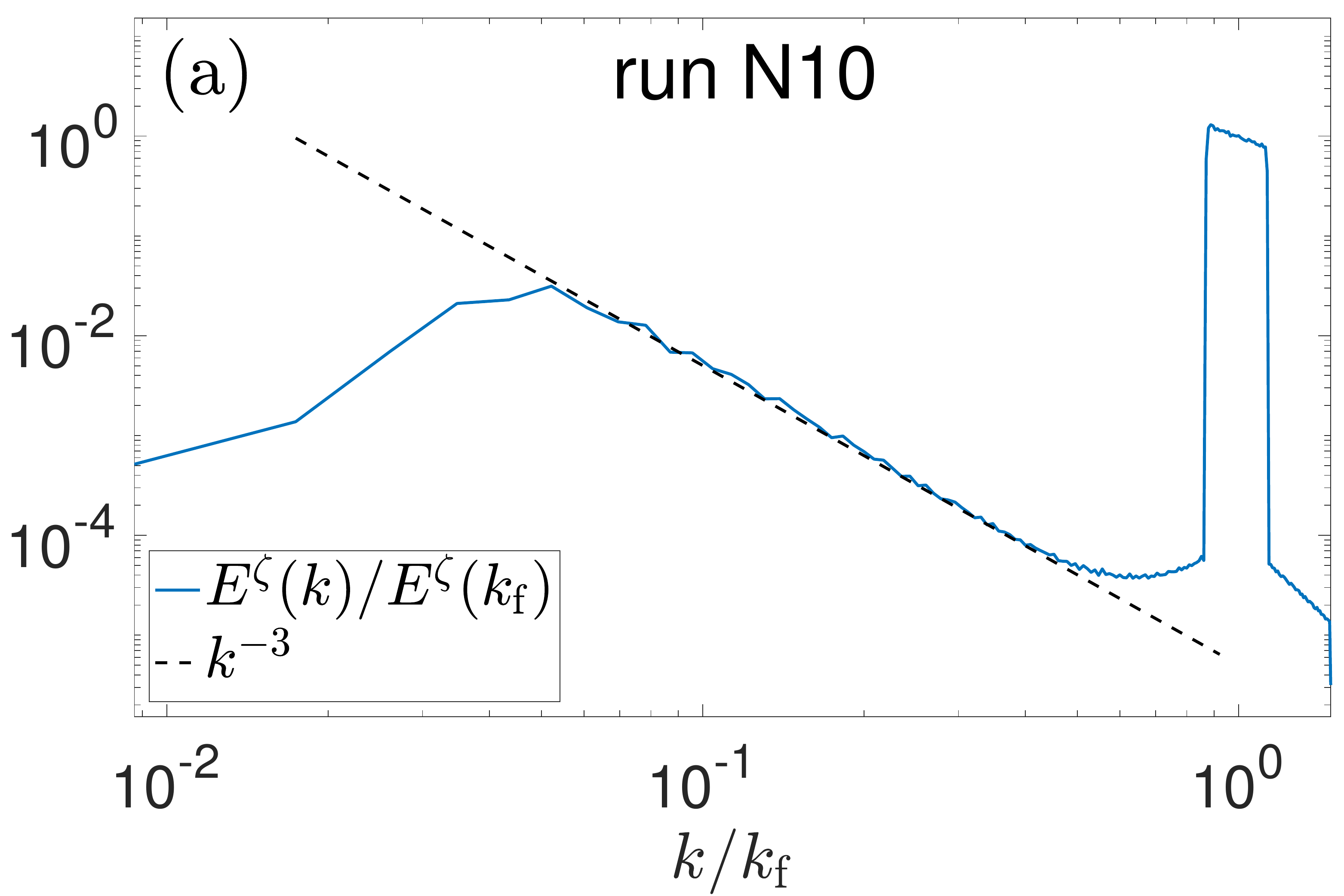}

\includegraphics[clip,width=8.5cm]{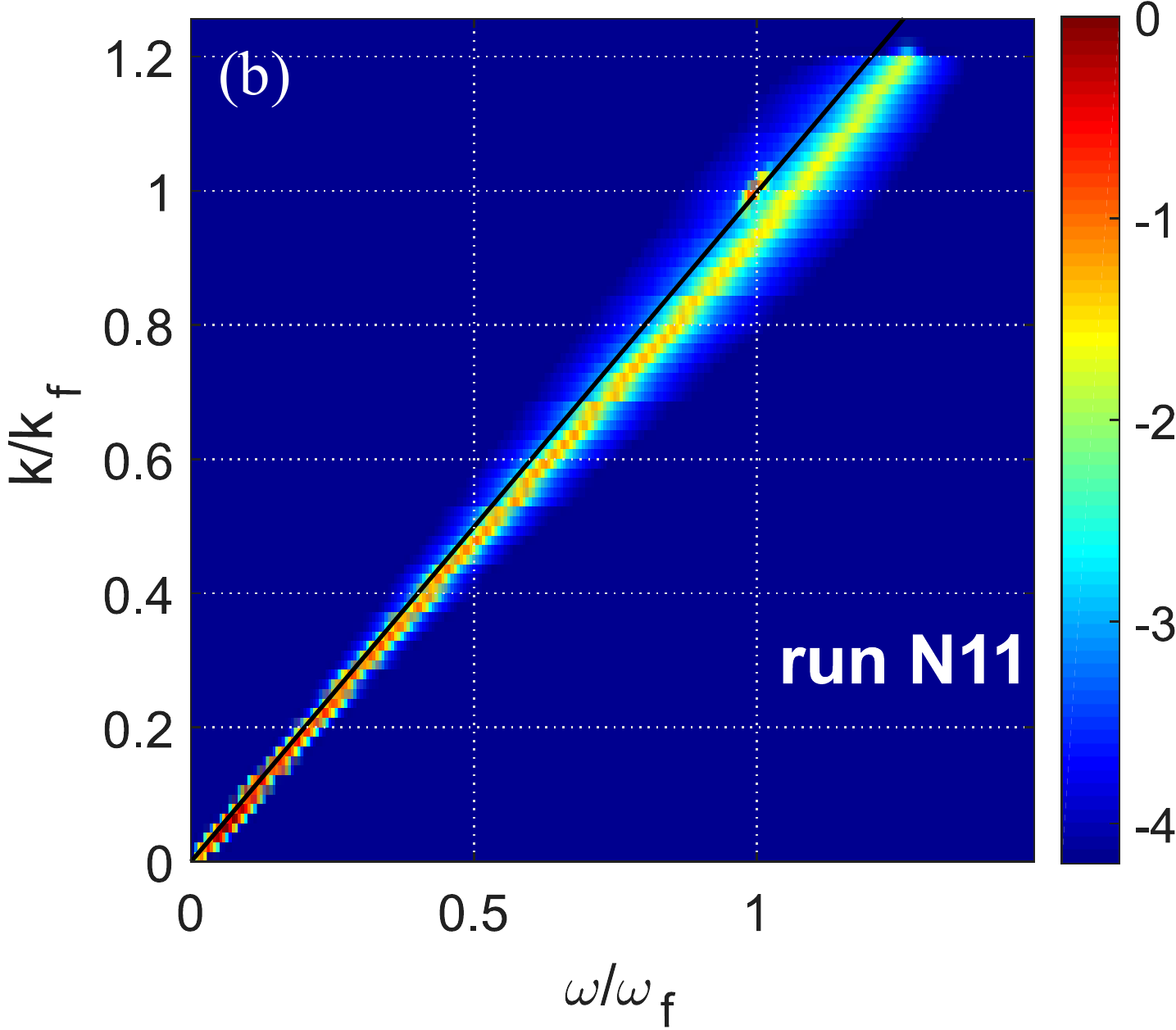}
\caption{(a) DNS N10 forced at small scale. The dashed line shows the $k^{-3}$ scaling expected from the WTT predictions. (b) Wavenumber-frequency spectrum of the deformation for a moderate resolution simulation N11 forced at $k_f=200$~rad/m. The solid black line is the linear dispersion relation.}
\label{st_inverse_membrane}
\end{figure}
To check the presence of an inverse cascade, we perform numerical simulations by forcing at high wavenumbers. As shown in fig.~\ref{st_inverse_membrane}(a) the high resolution DNS N10 shows the development of an inertial range at scales larger than the forcing one. The scaling observed in the inertial range is fully compatible with the WTT prediction for the inverse cascade of wave action. Figure~\ref{st_inverse_membrane}(b) shows the frequency-wavenumber spectrum of a moderate resolution DNS N11. Again, the energy is localized in the vicinity of the linear dispersion relation in agreement with the phenomenology of WTT.

\subsection{Evolution of the non-linearity with the scale}
\begin{figure}[!htb]
  \centering
   \includegraphics[width=7.5cm]{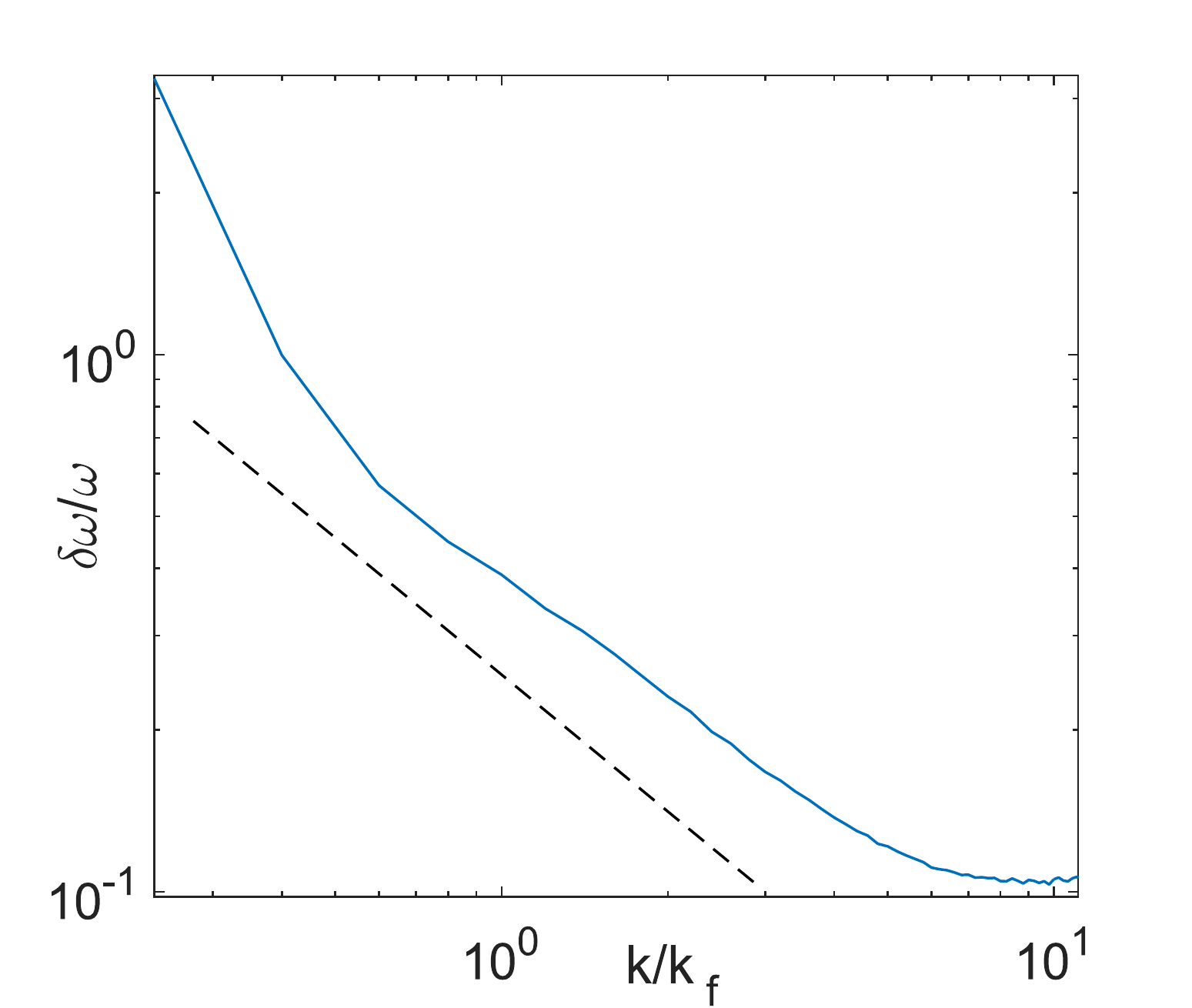}
  \caption{Evolution of the normalized spectral widening $\delta \omega/\omega$ as a function of k for the pure membrane (DNS N8). The dashed line is the WTT prediction $\delta\omega/\omega\propto k^{-2/3}$.} 
 \label{nonlinear}
\end{figure}

The hypothesis underlying the WTT is the scale separation between the slow nonlinear time scale $t_{NL}$ of the evolution of the wave spectrum and the fast oscillation of the linear time scale $t_L=2\pi/\omega_k$. Following Newell et al.~\cite{NNB}, in the framework of WTT, the ratio $t_{NL}/t_{L}$ is supposed to behave as $t_{NL}/t_{L}\propto P^{-2/3}k^{-2\gamma/3+2\alpha}$ where $\gamma$ is the degree of homogeneity in $k$ of the scattering matrix $J^{-1 s_1 s_2 s_3}_{-{\bf p} {\bf k}_1{\bf k}_2{\bf k}_3} $ and $\alpha$ is the degree of homogeneity of the dispersion relation. Here $\gamma=2$ and $\alpha=1$ for the membrane so that:
\begin{eqnarray}
t_{NL}/t_{L}\propto& \, P^{-2/3}k^{2/3} 
\end{eqnarray}
It means that the nonlinearity is getting weaker with $k$. Thus no breakdown of the weak nonlinearity hypothesis should be expected at high $k$ if the nonlinearity is weak at low $k$.
The nonlinear time scale appears as a widening $\delta \omega_{NL}\propto 1/t_{NL}$ of the energy concentration around the dispersion relation. Thus the ratio $\delta \omega_{NL}/\omega_k$ should behave as
\begin{eqnarray}
\delta \omega_{NL}/\omega_k\propto &\, P^{2/3}k^{-2/3} 
\end{eqnarray}

The widening $\delta \omega$ can be extracted from the wavenumber-frequency spectrum $E(k,\omega)$. At a given value of $k$, a Gaussian profile in $\omega$ can be fitted to $E(k,\omega)$ (of fig.~\ref{st_weak}(b)) around the maximum of energy (which corresponds to the dispersion relation). The result is shown in fig.~\ref{nonlinear}. The observed decay for the spectral width of the membrane is indeed following the scaling $1/k^{2/3}$ which supports the fact that the observed turbulence in the membrane is consistent with the WTT.

\section{Conclusion}

In this article, we compared experimentally and numerically the statistical properties of wave turbulence in a vibrating membrane to that of weak turbulence in a vibrating plate that had been studied before. The fundamental difference between the membrane and the plate is that the waves are not dispersive in the membrane as opposed to the case of the plate. Other systems such as acoustics or shallow water surface waves show that weak turbulence is no longer present but coherent structures such as shocks or solitons are observed. Here no such structures are observed when evolving continuously from the plate to the membrane and a state of weak turbulence is observed numerically in the membrane. This comes from the fact that the nonlinear coupling vanishes for collinear waves so that to prevent a strong cumulative effect of nonlinearity along rays yielding shocks for instance. Here only $2\leftrightarrow2$ interactions are possible that cause an angular redistribution of energy so that the systems remains weakly turbulent. A difference between the membrane and the plate is thus the presence of an inverse cascade for the membrane that should not exist for the plate (which is actually the object of a controversy~\cite{During:2015gt}). Thus there is a class of such systems that can develop weak turbulence of non dispersive waves. The case of weak gravitational waves is similar, the nonlinear coupling also vanishes for $3\leftrightarrow1$ interactions~\cite{Galtier:2017hi} so that weak turbulence should also exist for such waves.

\appendix

\section{Saddle-point like approximation}
\label{Saddle-node}
Wave turbulence theory is based on a multi-scale perturbation expansion, in which the kinetic equation emerges as a condition to keep bounded the long time limit behavior of the expansion. The general formalism can be found in several works \cite{Saffman,NNB}, in particular for elastic plates in \cite{platesPhysD}. The wave turbulence formalism requires the understanding of the long time behavior of the integrals 

\begin{equation}
I({\bf p})= \int F({\bf k}_1,{\bf k}_2,{\bf p}) \frac{e^{ i t \, h({\bf k}_1,{\bf k}_2,{\bf p}) }  -1}{i  h({\bf k}_1,{\bf k}_2,{\bf p} )}\, d{\bf k}_1d{\bf k}_2,
\label{resonh1}
\end{equation}
and 
\begin{eqnarray} 
E({\bf p})&=& \int F({\bf k}_1,{\bf k}_2,{\bf p}) \int_0^t \frac{e^{ i \tau \, h({\bf k}_1,{\bf k}_2,{\bf p}) }  -1}{i  h({\bf k}_1,{\bf k}_2,{\bf p} )} d\tau\, d{\bf k}_2d{\bf k}_1\nonumber\\
\label{sec1}
\end{eqnarray}
where $F({\bf k}_1,{\bf k}_2,{\bf p})$ is a regular function and as a shorthand notation one defines  $$h({\bf k}_1,{\bf k}_2,{\bf p})=s_1\omega({\bf k}_1)+s_2\omega({\bf k}_2)+s_3\omega({\bf k}_3)+s\omega({\bf p})$$ with  ${\bf k}_3={\bf p}-{\bf k}_1-{\bf k}_2$. In general for non dispersive waves the later expression $I({\bf p})$ and $E({\bf p})$ has a well established asymptotic behaviors given by  
\begin{widetext}
\begin{eqnarray}
\lim_{t\rightarrow \infty } I({\bf p})=\int \text{sign}(t)\pi F({\bf k}_1,{\bf k}_2,{\bf p})\delta(h({\bf k}_1,{\bf k}_2,{\bf p}))\,  d{\bf k}_2  d{\bf k}_1+iP\int  \frac{F({\bf k}_1,{\bf k}_2,{\bf p})}{ h({\bf k}_1,{\bf k}_2,{\bf p} )}\, d{\bf k}_2  d{\bf k}_1\nonumber\\
\label{resonh5}
\end{eqnarray}
and 
\begin{eqnarray}
\lim_{t\rightarrow \infty } \frac{E({\bf p})}{t} &=& \int \text{sign}(t)\pi F({\bf k}_1,{\bf k}_2,{\bf p})\delta(h({\bf k}_1,{\bf k}_2,{\bf p}))   d{\bf k}_2d{\bf k}_1+iP\int \frac{F({\bf k}_1,{\bf k}_2,{\bf p}) }{h({\bf k}_1,{\bf k}_2,{\bf p} )} d{\bf k}_2d{\bf k}_1
\label{LTBE}
\end{eqnarray}
\end{widetext}
with $P$ the principal value of the integral.  A good comprehension of these asymptotic behaviors are crucial for  studying the non dispersive limit. Therefore in this Appendix we give a detailed derivation of the asymptotic behaviors (\ref{resonh5}) and (\ref{LTBE}).

The  long time limit of the integral (\ref{resonh1})  is dominated by the behavior near the manifold defined by  $h({\bf k}_1,{\bf k}_2,{\bf p})=0$, namely the resonant manifold. If no resonant manifold exist i.e. $h({\bf k}_1,{\bf k}_2,{\bf p})\neq 0$  the Riemann-Lebesgue lemma implies  that the oscillatory term vanish in the limit $t\rightarrow\infty$, hence 
\begin{equation}
\lim_{t\rightarrow\infty}I({\bf p})= i\int \frac{F({\bf k}_1,{\bf k}_2,{\bf p}) }{  h({\bf k}_1,{\bf k}_2,{\bf p} )}\, d{\bf k}_1d{\bf k}_2.
\label{resonh2}
\end{equation}
and no evolution of the wave amplitude exists since  only the real part of $I({\bf p})$ contributes to it.
If a resonant manifold exists the integral  $I({\bf p})$ is dominated by the region close to such manifold. Generically one can consider that for each ${\bf p}$ and ${\bf k}_1$ exist a vector  ${\bf k}^*_2(\alpha)$   parametrized by a single parameter $\alpha$ which satisfies $h({\bf k}_1,{\bf k}^*_2,{\bf p} )=0$ ( we will omit unless necessary  $\alpha$ in ${\bf k}^*_2$) . 
 Considering that the resonant manifold is contained in a small $2d$ dimensional volume $\Omega$ the integral $I({\bf p})$ can be rewritten as
\begin{eqnarray}
\lim_{t\rightarrow\infty}I({\bf p})&=&\lim_{t\rightarrow \infty }\int_\Omega F({\bf k}_1,{\bf k}_2,{\bf p}) \frac{e^{ i t \, h({\bf k}_1,{\bf k}_2,{\bf p}) }  -1}{i  h({\bf k}_1,{\bf k}_2,{\bf p} )}\, d{\bf k}_2d{\bf k}_1\nonumber\\
&+&i\int_{\mathbb{R}^{2d}/\Omega} \frac{F({\bf k}_1,{\bf k}_2,{\bf p}) }{  h({\bf k}_1,{\bf k}_2,{\bf p} )}\, d{\bf k}_2d{\bf k}_1,
\label{resonh2a}
\end{eqnarray}
where in the region $\mathbb{R}^{2d}/\Omega$ we made use of the Riemman-Lebesgue lemma.
Close to the resonant manifold (i.e. ${\bf k}_2 \in \Omega$)

\begin{equation}
h({\bf k}_1,{\bf k}_2,{\bf p})\approx \nabla_{{\bf k}_2}h({\bf k}_1,{\bf k}^*_2,{\bf p})\cdot\delta {\bf k}_2+\ldots
\label{app}
\end{equation}

 with $$\nabla_{{\bf k}_2}h({\bf k}_1,{\bf k}^*_2,{\bf p})=s_2\omega'(k^*_2)\frac{\bf k^*_2}{|k^*_2|}-s_3\omega'(k^*_3)\frac{\bf k^*_3}{|k^*_3|}$$   and $\delta {\bf k}_2={\bf k}_2-{\bf k}^*_2$. If the gradient $\nabla_{{\bf k}_2}h({\bf k}_1,{\bf k}^*_2,{\bf p})$ vanishes at the resonant manifold, the approximation (\ref{app}) is not correct and the asymptotic behaviors (\ref{LTBE}) and (\ref{resonh5}) are not longer valid. Such is the case for most non dispersive systems, and is the main cause for the breakdown of the standard wave turbulence theory on these systems. The particular case of an elastic membrane will be discussed in Appendix \ref{non-dispersive}. 
Using one more time the Riemman-Lebesgue lemma in (\ref{resonh2a}) in the limit $\lim{t\rightarrow\infty} $ one gets
\begin{equation}
 I({\bf p})= \int F({\bf k}_1,{\bf k}_2,{\bf p}) \frac{e^{ i t \, \nabla_{{\bf k}_2}h({\bf k}_1,{\bf k}^*_2,{\bf p})\cdot\delta {\bf k}_2 }  -1}{i  h({\bf k}_1,{\bf k}_2,{\bf p} )}\, d{\bf k}_2d{\bf k}_1.
\label{resonh3}
\end{equation}

One can easily corroborate that $\nabla_{{\bf k}_2} h({\bf k}_1,{\bf k}^*_2,{\bf p})\cdot{\bf k}_2^\parallel=\partial_\alpha h({\bf k}_1,{\bf k}^*_2,{\bf p})=0$ where ${\bf k}_2^\parallel\equiv \partial_\alpha {\bf k}^*_2$ is the tangent vector of the resonant manifold. Therefore, defining  the orthogonal vector ${\bf k}_2^\perp$ such $ \nabla_{{\bf k}_2}h({\bf k}_1,{\bf k}^*_2,{\bf p})\cdot{\bf k}_2^\perp=||\nabla_{{\bf k}_2}h({\bf k}_1,{\bf k}^*_2,{\bf p})|| k_2^\perp $, one can perform a change of variable which  leads to 
\begin{multline}
\lim_{t\rightarrow\infty} I({\bf p})=\lim_{t\rightarrow \infty }\int G({\bf k}_1,{\bf k}_2,{\bf p}) \\
\times\frac{e^{ i t \, ||\nabla_{{\bf k}_2}h({\bf k}_1,{\bf k}^*_2,{\bf p})|| k_2^\perp }  -1}{i  h({\bf k}_1,{\bf k}_2,{\bf p} )}\, dk^\perp_2 d\alpha  d{\bf k}_1
\label{resonh4}
\end{multline}
where $G({\bf k}_1,{\bf k}_2,{\bf p})=F({\bf k}_1,{\bf k}_2,{\bf p})\mathcal{J}({\bf k}_1,{\bf k}_2,{\bf p} )$ and $\mathcal{J}({\bf k}_1,{\bf k}_2,{\bf p} )$ is the determinant of the Jacobian of the change of variable. We shall consider that $G({\bf k}_1,{\bf k}_2,{\bf p})$ is a regular function.

The integral in the variable $dk^\perp_2$ is of the type $$\tilde{I}=\int^{\infty}_{-\infty}g(x)\frac{e^{i h'(0) x t }-1}{i h(x) } dx,$$ where $h(x)\approx h'(0)x$ for $|x|\approx0$ and $h'(0)>0$. The integrand is regular in $0$ hence  
\begin{eqnarray}
\tilde{I}&=&\lim_{\epsilon\rightarrow0}\left(\int^{-\epsilon}_{-\infty}g(x)\frac{e^{i h'(0) x t }}{i h(x) } dx+\int^{\infty}_{\epsilon}g(x)\frac{e^{i h'(0) x t }}{i h(x) } dx\right)\nonumber\\
&+&iP\int^{\infty}_{-\infty}\frac{g(x)}{h(x)}dx
\end{eqnarray}
and a simple contour integrations leads to 
\begin{equation}
\tilde{I}=\text{sign}(t)\pi \frac{g(0)}{h'(0)}+iP\int^{\infty}_{-\infty}\frac{g(x)}{h(x)}dx.
\end{equation}
Therefore the expression (\ref{resonh4})  is given by 
\begin{widetext}
\begin{eqnarray}
\lim_{t\rightarrow\infty} I({\bf p})&=&\int \text{sign}(t)\pi \frac{G({\bf k}_1,{\bf k}^*_2,{\bf p})}{  ||\nabla_{{\bf k}_2}h({\bf k}_1,{\bf k}^*_2,{\bf p})||}\, \delta(k^\perp_2)\,  dk^\perp_2 d\alpha  d{\bf k}_1+iP\int  \frac{G({\bf k}_1,{\bf k}_2,{\bf p})}{  h({\bf k}_1,{\bf k}_2,{\bf p} )}\, dk^\perp_2 d\alpha  d{\bf k}_1\nonumber\\
\label{resonh6}
\end{eqnarray}
The first term is evaluated at $k^\perp_2=0$ which correspond to the resonant manifold given by the solutions of  $h({\bf k}_1,{\bf k}_2,{\bf p})=0$, hence we can rewrite the last expression as
\begin{eqnarray}
\lim_{t\rightarrow\infty} I({\bf p})&=&\int \text{sign}(t)\pi G({\bf k}_1,{\bf k}_2,{\bf p})\delta(h({\bf k}_1,{\bf k}_2,{\bf p}))\,  dk^\perp_2d\alpha  d{\bf k}_1+iP\int  \frac{G({\bf k}_1,{\bf k}_2,{\bf p})}{  h({\bf k}_1,{\bf k}_2,{\bf p} )}\, dk^\perp_2 d\alpha  d{\bf k}_1.
\label{resonh7}
\end{eqnarray}
Finally comming back to the original variables $dk^\perp_2 d\alpha\rightarrow d{\bf k}_2 $ one leads to the desired result (\ref{resonh5}).

The second class of integral (\ref{LTBE}) can be rewritten as 
\begin{eqnarray} 
E({\bf p})&=& \int F({\bf k}_1,{\bf k}_2,{\bf p})  \frac{1- e^{ i t \, h({\bf k}_1,{\bf k}_2,{\bf p}) }  + i t h({\bf k}_1,{\bf k}_2,{\bf p} )}{  h({\bf k}_1,{\bf k}_2,{\bf p} )^2} d\tau\, d{\bf k}_2d{\bf k}_1.
\label{sec11}
\end{eqnarray}
The same procedure  that was used to obtain (\ref{resonh4}) leads in the long time limit to the expression
\begin{eqnarray}
\lim_{t\rightarrow\infty} \frac{E({\bf p})}{t}&=&\lim_{t\rightarrow\infty}\frac{1}{t} \int G({\bf k}_1,{\bf k}_2,{\bf p}) \frac{1-e^{ i t \, ||\nabla_{{\bf k}_2}h({\bf k}_1,{\bf k}^*_2,{\bf p})|| k_2^\perp }  + i t h({\bf k}_1,{\bf k}_2,{\bf p} )}{  h({\bf k}_1,{\bf k}_2,{\bf p} )^2}\, dk^\perp_2 d\alpha  d{\bf k}_1\nonumber\\
\label{sec2}
\end{eqnarray}
The integral in the variable $dk^\perp_2$ is of the type $$\tilde{E}=\int^{\infty}_{-\infty}g(x)\frac{1-e^{i h'(0) x t }+i t h(x)}{ h(x)^2 } dx,$$ where $h(x)\approx h'(0)x$ for $|x|\approx0$ and $h'(0)>0$. Integrating by part once and considering $\partial_x h\neq 0$ one gets  
\begin{eqnarray}
\tilde{E}&=&-i\int^{\infty}_{-\infty}\frac{\partial}{\partial_x}\left(\frac{g(x)}{\partial_x h}\right)\frac{e^{i h'(0) x t }-1}{ ih(x) } dx+t \int^{\infty}_{-\infty}g(x)\frac{\frac{h'(0)}{\partial_x h}e^{i h'(0) x t }-1}{ ih(x) } dx
\end{eqnarray}
and a simple contour integrations leads to 
\begin{eqnarray}
\tilde{E}=t\left(\text{sign}(t)\pi \frac{g(0)}{h'(0)}+iP\int^{\infty}_{-\infty}\frac{g(x)}{h(x)}dx\right)-i\text{sign}(t)\pi \frac{1}{h'(0)}\frac{\partial}{\partial_x}\left(\frac{g(x)}{\partial_x h}\right)_{x=0}+P\int^{\infty}_{-\infty}\frac{1}{h(x)}\frac{\partial}{\partial_x}\left(\frac{g(x)}{\partial_x h}\right)dx.\nonumber\\
\end{eqnarray}
Therefore the expression (\ref{sec2})  is given by 
\begin{eqnarray}
\lim_{t\rightarrow\infty} \frac{E({\bf p})}{t}&=&\int \text{sign}(t)\pi \frac{G({\bf k}_1,{\bf k}^*_2,{\bf p})}{  ||\nabla_{{\bf k}_2}h({\bf k}_1,{\bf k}^*_2,{\bf p})||}\, \delta(k^\perp_2)\,  dk^\perp_2 d\alpha  d{\bf k}_1+iP\int  \frac{G({\bf k}_1,{\bf k}_2,{\bf p})}{  h({\bf k}_1,{\bf k}_2,{\bf p} )}\, dk^\perp_2 d\alpha  d{\bf k}_1\nonumber\\
\label{sec3}
\end{eqnarray}
and following the same procedure as for $I({\bf p})$ we obtain the desired result (\ref{LTBE}).

\end{widetext}

\section{Non-dispersive waves }
\label{non-dispersive}

For dispersive waves the long time behavior of the integrals (\ref{resonh5}) and (\ref{LTBE}) are guaranteed since the gradients $|\nabla_{{\bf k}_i}h({\bf k}_1,{\bf k}^*_2,{\bf p})|$ are non zero over the resonant manifold. 
For non dispersive waves the situation is more subtle. In particular, for acoustic waves \cite{aucoin,acustictur},  where the non-linearity is quadratic,  the local behavior of the associated function $h({\bf k}_1,{\bf k}_2,{\bf p})$ near the resonant manifold is not linear but quadratic. Indeed, it is easy to show that the gradients of $h({\bf k}_1,{\bf k}_2,{\bf p})$ are exactly zero on the manifold. In such cases the long time limit behavior of the integral strongly depends on the space dimension. It has been shown that for the $3$ dimensional case, it is possible to find a similar kinetic equation, but with a restriction that only allows resonant interactions between collinear waves. Therefore no angular energy redistribution is possible. 

The elastic plate under a strong tension corresponds to a two-space dimensional system with linear dispersion relation $\omega({\bf k})= \sqrt{T/\rho h} |{\bf k}|$ and a cubic non-linearity. The resonant manifold is given by
 \begin{equation}
h({\bf k}_1,{\bf k}_2,{\bf p})=c(s_1|{\bf k}_1|+s_2|{\bf k}_2|+s_3|{\bf p}-{\bf k}_1-{\bf k}_2|-|{\bf p}|)=0.
\label{manih}
\end{equation}
where $c=\sqrt{T/\rho h}$.
For the sake of simplicity we have chosen $s=-1$. Two different solutions (or manifolds) exist. The first solution is similar to the one found for the acoustic problem, which represents a collinear interaction, that is ${\bf k}^*_1=\alpha_1{\bf p}$ and ${\bf k}^*_2=\alpha_2{\bf p}$. Then, the resonant condition reduces to:
\begin{equation}
s_1|{\alpha_1}|+s_2|{\alpha_2}|+s_3|{1-\alpha_1-\alpha_2}|-1=0
\label{maniha}
\end{equation}
For the $3\leftrightarrow1$ interaction ($s_1=s_2=s_3=1$) one has that  the manifold is given  for any $(\alpha_1,\alpha_2)$ with $\alpha_1>0,\alpha_2>0$ and $1-\alpha_1-\alpha_2>0$. For the $2\leftrightarrow2$ interaction ($s_1=s_2=1$ and $s_3=-1$) the solution is given for any $(\alpha_1,\alpha_2)$ while  $\alpha_1>0,\alpha_2>0$ and  $\alpha_1+\alpha_2-1>0$.
 It is easy to see that near the resonant manifold the first contribution to $h({\bf k}_1,{\bf k}_2,{\bf p})$ is quadratic in ${\bf k}$ because  
$$\nabla_{{\bf k}_i}h({\bf k}^*_1,{\bf k}^*_2,{\bf p})=c\frac{\bf p}{|p|}\left(s_i\frac{\alpha_i}{|\alpha_i|}-s_3\frac{1-\alpha_1-\alpha_2}{|1-\alpha_1-\alpha_2|}\right)=0$$
 Then, the long time limit of integrals (\ref{resonh1}) and (\ref{sec1}) are not given by the expressions (\ref{resonh5}) and (\ref{LTBE}). In general the asymptotic behavior of (\ref{resonh1}) and (\ref{sec1}) is not well established, but fortunately for the particular case of an elastic plates the scattering coefficient   \eqref{Eq:ScatMat} strongly vanishes for collinear wave vectors.  Therefore, at least up to the second order expansion, the collisional integral is trivial.

\begin{figure}[h]
\centering
\includegraphics[width=.48\linewidth]{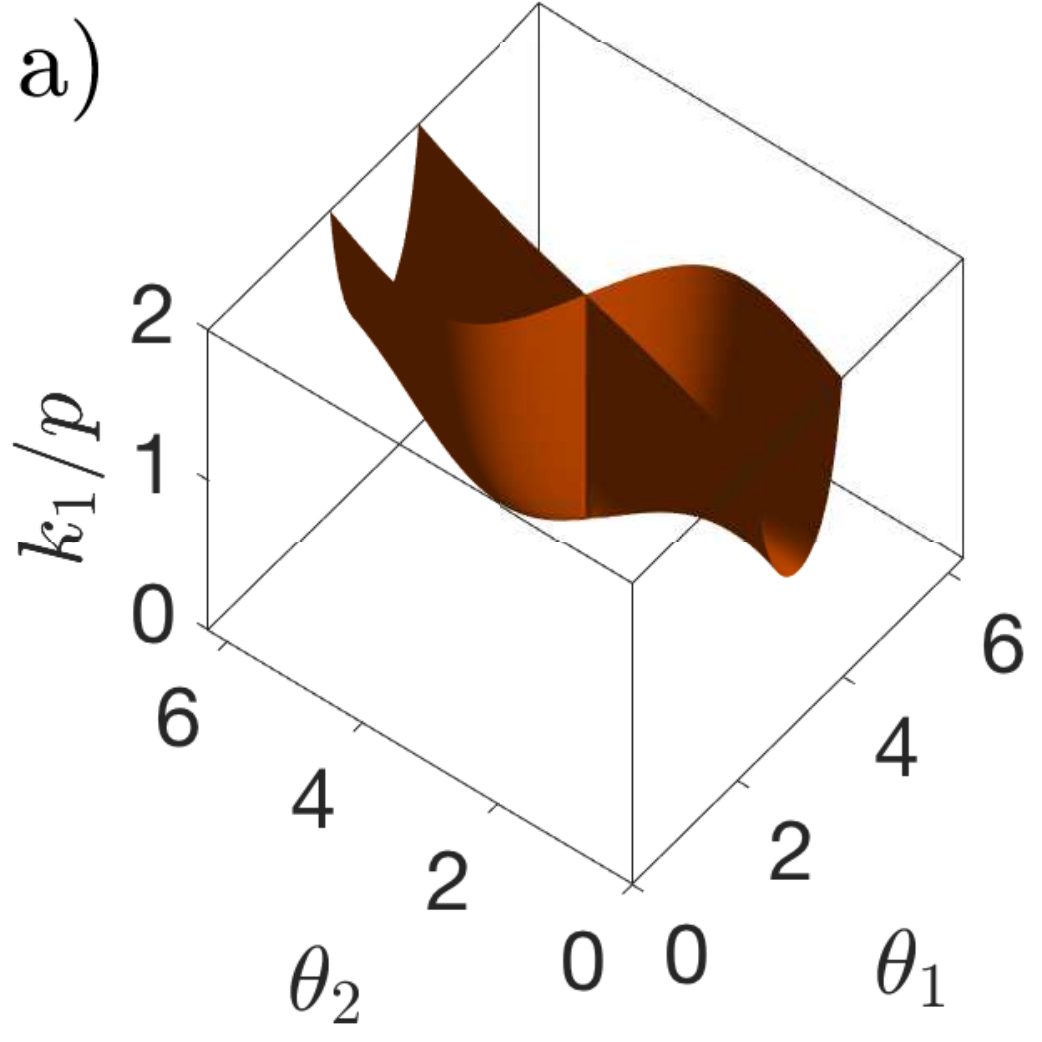}
\includegraphics[width=.48\linewidth]{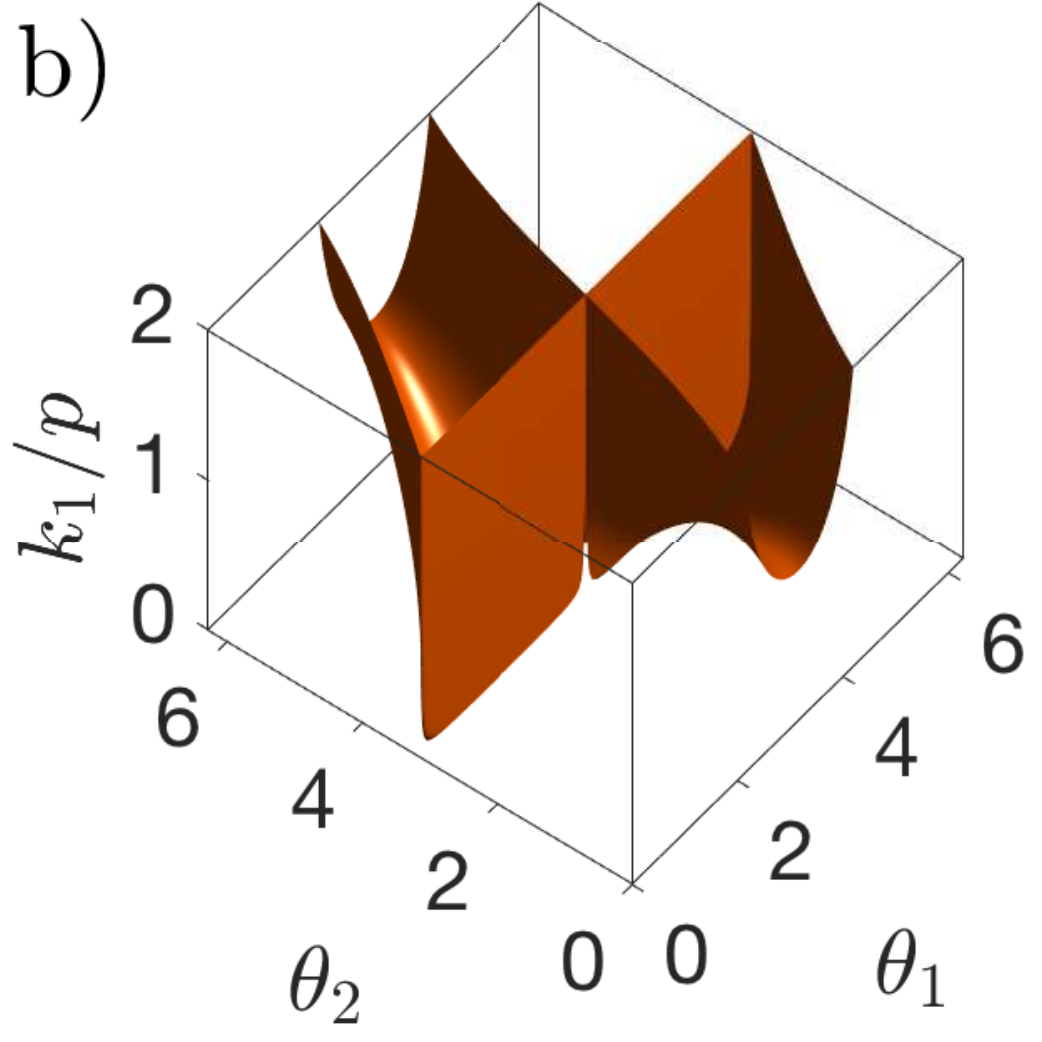}
\includegraphics[width=.48\linewidth]{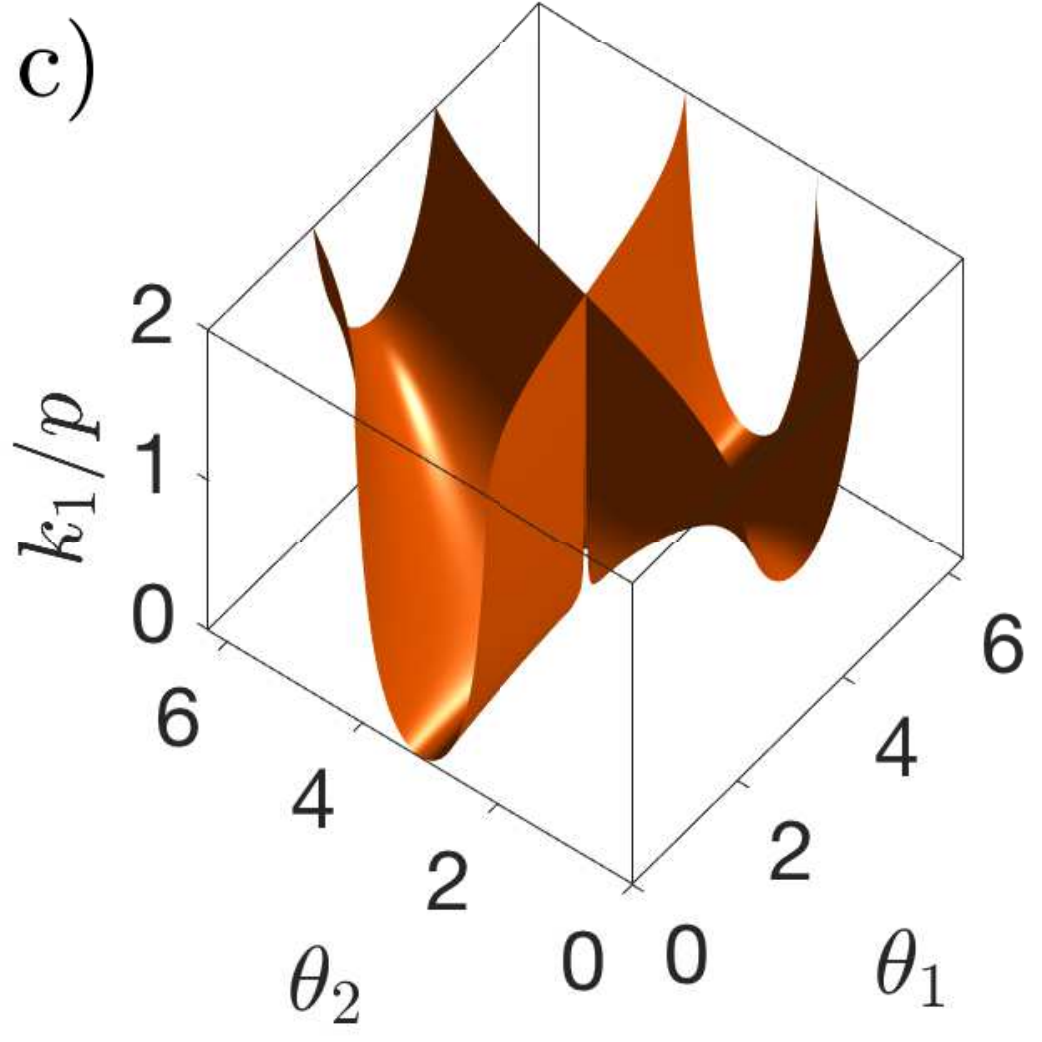}
\includegraphics[width=.48\linewidth]{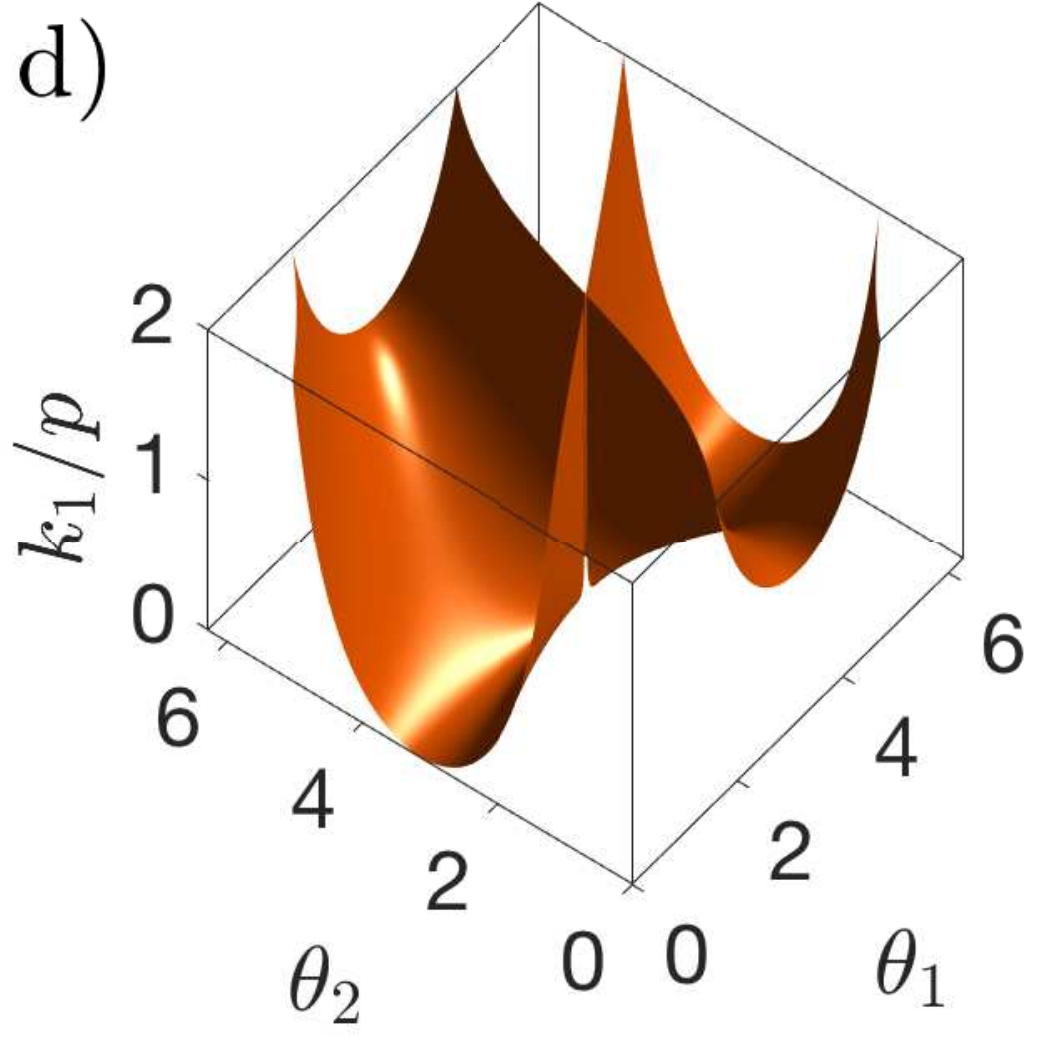}
\caption{The resonant manifold for $k_2/p=.5$ (a), $k_2/p=1$ (b), $k_2/p=1.5$ (a) and $k_2/p=100$ (b). $\theta_p=\pi$ by choosing the system of coordinates. }
\label{cone}
\end{figure}

Remarkably, as we said at the beginning, there exist another solution to (\ref{manih}) which corresponds to   $2\leftrightarrow2$ interaction of non-collinear waves.  This solution can be expressed in polar coordinates as:
\begin{eqnarray}
&&k_2=\frac{k_1p(1-\cos(\theta_p-\theta_1))}{k_1(1-\cos(\theta_2-\theta_1))-p(1-\cos(\theta_2-\theta_p))}\\
&&k_1+k_2>p.
\end{eqnarray}
The resonant manifold is displayed in Fig.\ref{cone} for different values of $k_2/p$. We have set $\theta_p=\pi$ by choosing the system of reference. The trivial collinear resonant manifold is given by curve $\theta_1=\theta_2=\pi$. 
One can easily verify that over the manifold described by such solution the gradient of $h({\bf k}_1,{\bf k}_2,{\bf p})$ is not zero and the standard weak wave turbulence description can be applied.  Finally one gets the same kinetic equation (\ref{kinetic1}) keeping in mind that the $3\leftrightarrow1$ interaction does not contribute. 

\begin{acknowledgements}
This project has received funding from the European Research Council (ERC) under the European Union's Horizon 2020 research and innovation programme (grant agreement No 647018-WATU).  GD  acknowledge support from FONDECYT grant N$^\circ$ 1181382 and N$^\circ$  1150463. We thank V. Govart, M. Kusulja and J. Virone for their technical assistance with the experiment. 
\end{acknowledgements}

\bibliography{mabiblio}

\end{document}